\documentclass{aa}

\usepackage{txfonts}
\usepackage{graphicx}
\usepackage[below]{placeins}
\usepackage{natbib}
\bibpunct{(}{)}{;}{a}{}{,}

\begin{document}

\title{Chemical evolution in \mbox{S\'ersic 159-03} observed with XMM-Newton}

\author{J. de Plaa \inst{1,2} \and
	N. Werner \inst{1} \and
        A. M. Bykov \inst{3}\and
	J. S. Kaastra \inst{1} \and
	M. M\'endez \inst{1} \and
	J. Vink \inst{1,2} \and
	J. A. M. Bleeker \inst{1,2} \and
	M. Bonamente \inst{4} \and
	J. R. Peterson \inst{5}
	}

	\institute{SRON Netherlands Institute for Space Research, Sorbonnelaan 2, 3584 CA Utrecht, The Netherlands\\
	\email{j.s.kaastra@sron.nl}
	\and Astronomical Institute, Utrecht University, PO Box 80000, 3508 TA Utrecht, The Netherlands
	\and A.F.~Ioffe Institute for Physics and Technology, 26 Polytechnicheskaia, 194021 St. Petersburg, Russia
	\and Department of Physics, University of Alabama, Huntsville, AL 35899, USA
	\and KIPAC, Stanford University, PO Box 90450, MS 29, Stanford, CA 94039,USA
	}
												  
\offprints{J. de Plaa, \email{j.de.plaa@sron.nl}}
													  
\date{Received 20 July 2005 / Accepted 26 February 2006}

\abstract{Using a new long X-ray observation of the cluster of galaxies S\'ersic 159-03 with XMM-Newton, we derive radial 
temperature and abundance profiles using single- and multi-temperature models. The fits to the EPIC and RGS spectra 
prefer multi-temperature models especially in the core. The radial profiles of oxygen and iron measured with
EPIC/RGS and the line profiles in RGS suggest that there is a dip in the O/Fe ratio in the centre of the
cluster compared to its immediate surroundings. A possible explanation for the large scale metallicity distribution 
is that SNIa and SNII products are released in the ICM through ram-pressure stripping of in-falling galaxies. This causes 
a peaked metallicity distribution. In addition, SNIa in the central cD galaxy enrich mainly the centre of the cluster 
with iron. This excess of SNIa products is consistent with the low O/Fe ratio we detect in the centre of the cluster. 
We fit the abundances we obtain with yields from SNIa, SNII and Population-III stars to derive the clusters 
chemical evolution. We find that the measured abundance pattern does not require a Population-III star 
contribution. The relative contribution of the number of SNIa with respect to the total number of SNe which 
enrich the ICM is about 25--50\%. Furthermore, we discuss the possible presence 
of a non-thermal component in the EPIC spectra. A potential source of this non-thermal emission can be 
inverse-Compton scattering between Cosmic Microwave Background (CMB) photons and relativistic electrons, 
which are accelerated in bow shocks associated with ram-pressure stripping of in-falling galaxies.

}

\maketitle

\setlength{\tabcolsep}{1.5mm}

\section{Introduction}

Hot diffuse X-ray emitting gas dominates the visible mass in clusters of galaxies,
but the structure and evolution of the cluster is not yet fully understood. During its formation, 
supernova explosions and galactic winds of member galaxies have enriched the 
Intra-Cluster Medium (ICM) substantially \citep{deyoung1978}. The abundances can 
provide the relative contribution of Supernova type Ia (SNIa), Supernova type II (SNII) and 
population-III stars (PopIII) to the enrichment of 
the ICM \citep[e.g][]{iwamoto1999,tsujimoto1995}, because the abundance ratios of the 
various elements are signatures of supernova types Ia and II and 
possibly of the remains of PopIII stars \citep[e.g.][]{gibson1997,loewenstein2001,baumgartner2005}. 
The radial distribution of the metals provides information about dynamical ways
to enrich the ICM like, for example, ram-pressure stripping \citep[e.g.][]{schindler2005}.

Clusters of galaxies appear as knots in the cosmic web. They 
accrete gas from the surrounding filaments, which consist of warm gas with a temperature 
in the range of 10$^{5-6}$ K. According to numerical hydro-dynamical simulations by e.g. 
\citet{cen1999} and \citet{dave2001}, this Warm-Hot Intergalactic Medium (WHIM)
could contain about half of the missing baryons in the universe, 

There have been several attempts to detect this WHIM in emission.
In the late 90's \citet{lieu1996a} and \citet{mittaz1998} discovered  
a soft X-ray excess in EUV and ROSAT spectra of several clusters. More recent observations 
with XMM-Newton appear to confirm the presence of a soft excess in some cluster spectra 
\citep{kaastra2003a,finoguenov2003}. The detection of a possibly redshifted
O VII line, which traces gas with a temperature of $\sim$ 10$^6$ K, suggests
that the gas might be the WHIM. Unfortunately, current instruments do not have
sufficient spectral resolution to prove that the emission is indeed extragalactic.

Recent measurements of S\'ersic 159-03 by \citet{bonamente2005} and \citet{kaastra2003a} 
show that the soft-excess can be fit both using thermal and non-thermal models. Inverse-Compton 
scattering of CMB photons with relativistic electrons can also contribute a non-thermal 
power-law component to the spectra. This mechanism was already proposed by \citet{sarazin1998}
to explain the extreme-ultraviolet emission from clusters. In the hard X-ray band up to 80 keV, 
detections of non-thermal emission have been claimed in several clusters, for example \object{Coma} 
and Abell 2256 \citep{fusco-femiano1999,fusco-femiano2005}. But these BeppoSAX 
detections of the hard-excess are still subject to debate \citep{rossetti2004}.

The cluster of galaxies \object{S\'ersic 159-03}, also known as \object{ACO S 1101}, 
was discovered by \citet{sersic1974}. Since then it was studied 
in X-rays as part of several cluster samples: e.g. EXOSAT \citep{edge1991} and 
ROSAT \citep{allen1997}. \citet{kaastra2001} reported results from an XMM-Newton 
observation with a useful exposure time of about 35 ks showing a radial temperature 
profile which peaks at $kT$ = 2.7 keV at a radius of 2$\arcmin$ 
from the core. The temperature drop in the core is relatively modest, while the 
temperature outside the 2$\arcmin$ radius drops rapidly to values around 0.5 keV. 
Because \object{S\'ersic 159-03} is thought to show a large soft X-ray excess, the cluster
was also included in the sample of \citet{kaastra2003a}. 

In this paper we present results from a 121 ks long XMM-Newton \citep{jansen2001} 
observation of \object{S\'ersic 159-03}. The main goal of the paper is to obtain accurate
temperature and abundance profiles as far out from the core as possible using a new method 
for handling the background. We exploit the large effective area of XMM to obtain temperature 
and abundance profiles with EPIC \citep{turner2001} and RGS \citep{herder2001}. This 
deep observation also allows us to study the nature of the previously detected soft 
excess in more detail, and in addition it provides more accurate radial profiles of 
the temperature and metal abundances. We then fit the derived abundances to 
yields of Supernovae type Ia and II, and PopIII stars. Moreover, we 
discuss the potential presence of non-thermal emission in the cluster.

Throughout this paper we use H$_{0}$ = 70 km s$^{-1}$ Mpc$^{-1}$, $\Omega_{\mathrm{m}}$ = 0.3 and $\Omega_{\Lambda}$ = 0.7.
Using this cosmology 1$\arcmin$ is 73 kpc at the cluster redshift of 0.0564 \citep{maia1987}.    
The elemental abundances presented in this paper are given relative to the solar abundances 
from \citet{lodders2003}.

\section{Observations and data analysis}

The XMM-Newton observation of \object{S\'ersic 159-03} was performed on November 20, 2002 and had a 
total duration of 121 ks. The two EPIC MOS cameras were operated in Full Frame mode and the EPIC pn camera 
in Extended Full Frame mode. For all EPIC cameras the thin filter was used. The RGS instruments were 
operated in the standard spectroscopy mode.
All data were analysed with the 6.1.0 version of the XMM Science Analysis System (SAS).

\subsection{EPIC Analysis}

One of the most important things to account for in extended source analysis is the background. 
Because we intend to measure the cluster properties as far from the core of the cluster as possible,
we need an accurate estimate of the local background, especially for the dim outer parts of the cluster.
In general the source fills the entire field of view, which makes a direct measurement of the local 
background very difficult.
Common practice is to extract spectra from a combined event list of several observations of empty fields,
like the ones compiled by \citet{lumb2002} or \citet{read2003}, and use them as best estimate for the 
local background. The spectra extracted from this blank field are scaled and then subtracted from the
source spectra. This method works fine in areas in which the surface brightness of the source is high, 
and with clusters that have similar background conditions (e.g. particle background, local cosmic background 
and instrumental background) compared to the blank fields.

The Cosmic X-ray Background (CXB), however, consists of multiple components, which are
more extensively described in Sect.~\ref{sec:cxb}. The emission can be roughly divided in two parts: a soft 
thermal component originating from hot plasma in our own galaxy and a non-thermal power-law component
caused by unresolved distant point sources, predominantly AGN. The soft galactic CXB component 
varies spatially across the sky.  Therefore, the total photon background at low energies 
can vary up to 30--60\% from pointing to pointing \citep{read2003}.
It is clear that when the background conditions in the source observation are very different from the
average blank-sky background, this can lead to systematic uncertainties in the fitted parameters from 
the outer parts of the cluster. 

\begin{figure}[t]
\begin{center}
\includegraphics[width=\columnwidth]{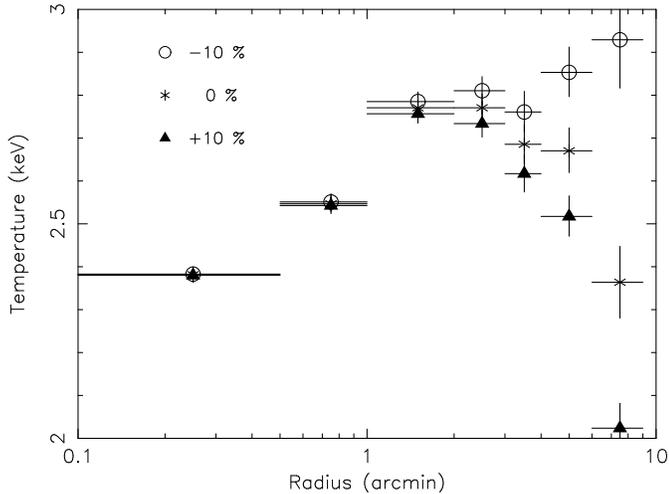}
\end{center}
\caption{Systematic effects in a temperature profile obtained with a blank-sky background dataset \citep{read2003}. 
The points with the cross are derived from spectra with no background scaling. The background for the circled 
data points is scaled with 0.9 and the triangles with 1.1.}
\label{fig:temp_sys}
\end{figure}

In Fig.~\ref{fig:temp_sys} we show the systematic effect of background subtraction on a temperature profile
extracted from the EPIC data of S\'ersic 159-03. For this plot we deliberately introduced an error of +10\% and -10\% 
in the background normalisation. The crosses show a profile derived with no background scaling at all,
while the circles and triangles show the temperature profile for a scaling of -10\% and +10\%, respectively.
From this plot we see that overestimating the background results in a lower temperature, and 
vice versa. Note that we included a 10\% systematic error on the background during 
fitting. We conclude that the uncertainties in the temperature are mainly caused by the background 
scaling and therefore we can gain a lot in accuracy by constraining the background normalisations.

In order to find the best estimate for the local background we divide the background into three components:
soft-protons, instrumental background, and the CXB. By separating
these components in our analysis, we will be able to constrain the normalisation and accuracy of each component.
The method we use is similar to the one used by \citet{deluca2004}, but with a few modifications. The steps 
are described below.

\subsubsection{Soft Protons}

In order to minimise the effect of soft protons in the spectra, we cut out the time intervals which show enhanced soft-proton flux
using the method described in \citet{pratt2002}.
We make a light curve using the events with energies between 2--12 keV and with a bin size of 100 s. From this light curve we extract 
the distribution of counts per bin. The resulting histogram is then fitted with a Poissonian function to determine the average number of counts per bin. 
The thresholds are subsequently fixed to $N \pm 2\sqrt{N}$, where $N$ is equal to the mean number of counts within a 100 s bin. This way we cut out 
every flare which reaches the 2$\sigma$ level. This method is more strict in rejecting soft-proton flares than using the conventional
$>$10 keV light curve with 3$\sigma$ clipping. Because of the broad energy interval we choose, the signal-to-noise in the light curve 
is very high and even small flares stand out from the quiescent level. By putting a strict 2$\sigma$ threshold we exclude also small
flares without losing a lot of usable data ($\sim 4$\% max. is lost in this case). This way we minimise the effect of soft-proton flares which can bias
our temperature determination like in Fig.~\ref{fig:temp_sys}. After applying the threshold we obtain an effective exposure time
of 81 ks for both MOS instruments and 60 ks for pn.

The disadvantage of this method is that we do not necessarily subtract the quiescent level of soft protons. We can get an indication of
the magnitude of the soft-proton count rate by using CCD9 of the RGS instrument. For this CCD the photon count rate from the source 
is very low, so the count rate is dominated by soft protons and instrumental noise. For our observation we find that the RGS CCD9 count rate
is 0.086 cnts/s in the quiescent periods. Compared to all the soft-proton background count rates measured during the
lifetime of XMM-Newton, this residual soft proton background is low \citep{gonzalez2004}.   
Therefore, the influence on our analysis should be very small.

\subsubsection{Instrumental background}

A good template for the instrumental background can be obtained from a closed-filter observation. For this analysis
we use a closed-filter observation which was performed after the cooling of the EPIC and RGS instruments in November 2002.
The properties of the observations are listed in Table~\ref{tab:closed}.    

\begin{table}[!tb]
\caption{The closed filter observations that we use for the EPIC background subtraction.}
\begin{center}
\begin{tabular}{llll}
\hline\hline
Instrument	& Obs. date	&	Obs. ID		& 	Exposure	\\
\hline
MOS		& 2003-04-06	&	0150390101	&	200 ks	\\
pn		& 2003-08-12	&	0160362801	&	35 ks   \\
\hline
\end{tabular}
\end{center}
\label{tab:closed}
\end{table}

The instrumental background contains roughly two components: fluorescence lines and a power law. A study of the EPIC-pn 
background by \citet{katayama2004} shows that the variation of the fluorescence lines is much lower than the variation in the
continuum. We use the power-law component to model the variable hard-particle background and the intrinsic instrumental noise. 
This background component is mainly caused by hard-particles which are able to reach the detector even when the filter wheel 
is in closed position. We especially choose a long closed-filter observation, not taken during the passage of the Earth radiation 
belts, to avoid getting a too large variation in the hard-particle flux with respect to our S\'ersic 159-03 observation. 

Because these variations can have a significant effect on fitted parameters (see Fig.~\ref{fig:temp_sys}), we would like to be able to modify the 
normalisation of the hard-particle power law by a few percent without changing the normalisations of the instrumental lines.
In order to find the scaling factors for this normalisation, we can use the events registered outside the field of view 
(out-of-FOV) of the EPIC instruments in both the closed-filter and source observations. These non-illuminated parts of the 
CCD chips provide unbiased data on the power-law normalisation, also when a bright source is present in the field of view. 

We start by determining the power-law index of the instrumental power law by fitting a closed-filter spectrum extracted from
the full field of view. We assume that this power-law index will be constant over the whole detector, since the observation does not suffer
from vignetting and PSF effects of the mirror. We fit the spectrum with just a spectral redistribution file and no effective area file.
From these fits we obtain a photon index ($\Gamma$) of 0.15 for MOS and 0.37 for pn. The MOS value is consistent with the value 
of $\sim$0.2 obtained in the same way by \citet{deluca2004} 

We extract the events labelled as out-of-FOV events (\verb+FLAG==#XMMEA_16+) to determine the normalisation of the 
instrumental background. To be sure that photons and soft-protons that scatter into the shielded out-of-FOV region of the detector
are not polluting the measurement, we only count events registered outside a radius of 15.4$\arcmin$ from the centre of 
the field-of-view. We divide the out-of-FOV count rates in our S\'ersic 159-03 observation with the closed-filter out-of-FOV count 
rates in the 8--12 keV band to obtain an instrumental background scaling factor ($c$). The $c$ values are 1.03 $\pm$ 0.02, 0.97 $\pm$ 0.02 
and 0.97 $\pm$ 0.06 for MOS1, MOS2 and pn respectively. In this case the correction on the power-law normalisation is 
only $\pm$ 3\% and within the statistical error, but we prefer to use the most likely value for $c$. Despite the fact that 
our values for $c$ are consistent with being 1, this is a useful exercise, because the normalisation of closed filter spectra 
is known to vary by about 15\% \citep{deluca2004}. Already a deviation of a few percent can lead to discrepancies in the 
temperature determination.  

We cannot simply multiply the
closed-filter spectra with our values for $c$, because then the instrumental lines would not be well subtracted. The only
component that we have to scale is the power-law component. Therefore, we add (or subtract) a power law with the 
same slope ($\Gamma$) to the original closed filter spectra ($S_{\mathrm{orig}}$). The normalisation of the power law ($n$) we add 
is derived as follows. We first solve this simple system for $S_{\mathrm{orig}}$ to approximate the scaled spectrum $S_{\mathrm{scal}}$:
\begin{eqnarray}
\left\{ \begin{array}{l}
S_{\mathrm{scal}} = c S_{\mathrm{orig}}  \\
S_{\mathrm{scal}} = n E^{-\Gamma} + S_{\mathrm{orig}} , 
\end{array} \right. 
\end{eqnarray}
and work out $n$:
\begin{equation}
n = \frac{(c-1)}{E^{-\Gamma}} S_{\mathrm{orig}} .
\label{eq:closed_norm}
\end{equation}
We can subsequently calculate $n$ for every energy bin of energy $E$. We exclude energy intervals affected by 
instrumental lines. The mean value of $n$ over all these bins is determined by fitting a Gaussian to a histogram 
of $n$. This value will be used to add or subtract the power law from the closed filter spectra. Finally, we subtract 
this scaled spectrum from the source spectra.

\subsubsection{Cosmic X-ray background}
\label{sec:cxb}

Instead of subtracting the CXB from the spectra, like it is done with blank-field data, we include the CXB 
components during fitting. Because the CXB can vary spatially across the sky we can more easily
adapt the flux of each component to the local conditions. Unlike the case of the instrumental 
background and soft-protons, all the photons enter through the XMM mirrors in the same way as the source photons.
Therefore, the response files (rmf) and effective area files (arf) must be applied and we must fit the CXB simultaneously
with the source spectra.

For the fitting we use the spectral components described by \citet{kuntz2000} and \citet{deluca2004}. 
For the thermal components in this paper we use MEKAL models, but 
the thermal components in \citet{kuntz2000} are fitted with Raymond-Smith models \citep{raymond1977}. 
There are, however, significant differences in ionisation balance and line strengths between MEKAL and Raymond-Smith, 
because the MEKAL code includes many more ions and up-to-date atomic constants. 
Hence, a temperature determined from a spectrum using the Raymond-Smith model is different from a temperature
determined using MEKAL for the same dataset. To obtain consistency, we choose to convert the model temperatures 
from \citet{kuntz2000} to MEKAL temperatures to match the measured spectra by \citet{kuntz2000} in our modelling. 
Unfortunately, this transformed \citet{kuntz2000} model does not fit a spectrum extracted from the \citet{read2003} 
blank-fields. Therefore, we empirically fit the \citet{read2003} blank-fields with a power law, a soft, and a hard thermal component, 
which turns out to be a good description of the data. Based on the best fit and the model by \citet{kuntz2000} we fix the 
temperatures and metallicities to the values listed in Table~\ref{tab:cxb}

In order to estimate the normalisations of the local background components around S\'ersic 159-03
we fit them to the outermost annulus (9--12$\arcmin$ from the core of the cluster) using the temperatures derived
from the \citet{read2003} blank-fields. We add an additional thermal component for the cluster emission. 
In our final fits for all annuli we fix the normalisations of the background components to the fitted values 
of the 9--12$\arcmin$ annulus, which are listed in Table~\ref{tab:cxb}. For comparison, the 2--10 keV absorbed 
integrated intensity of the power-law component is 2.26 $\times$ 10$^{-11}$ erg cm$^{-2}$ s$^{-1}$ deg$^{-2}$, 
which is consistent with the value of (2.24 $\pm$ 0.16) $\times$ 10$^{-11}$ erg cm$^{-2}$ s$^{-1}$ deg$^{-2}$
found by \citet{deluca2004}.

\begin{table}[!tb]
\caption{The CXB background components we use in the fits of S\'ersic 159-03. The integrated absorbed intensity 
over the 0.3--10 keV energy range was calculated using an $N_{\mathrm{H}}$ value of 1.79 $\times$ 10$^{20}$ cm$^{-2}$.
In the thermal components the abundances are set to solar values.}
\begin{center}
\begin{tabular}{lllll}
\hline\hline
Component	& kT		& $\Gamma$ 	& Z$_{\sun}$	& Integrated intensity \\
		& (keV)		& 		&		& (erg cm$^{-2}$ s$^{-1}$ deg$^{-2}$) \\
\hline
Soft thermal 	& 0.070		&		& 0.3		& 2.23 $\times$ 10$^{-12}$ \\
Hard thermal 	& 0.20		&		& 1.0		& 1.03 $\times$ 10$^{-11}$ \\
Power law	& 		& 1.41 		&		& 3.16 $\times$ 10$^{-11}$ \\
\hline
\end{tabular}
\end{center}
\label{tab:cxb}
\end{table}

In order to avoid large background fluctuations in our extraction regions we cut out bright point sources
which were identified by eye. The unresolved point sources are taken into account by the CXB power law.

\begin{table}
\caption{Boundaries of the annular extraction regions used in our EPIC analysis.
The annuli are centred on the cluster centre.}
\begin{center}
\begin{tabular}{ccc}
\hline\hline
Annulus	& Inner boundary ($\arcmin$) & Outer boundary ($\arcmin$) \\
\hline
1	& 0.0		& 0.5 \\
2 	& 0.5		& 1.0 \\
3	& 1.0		& 2.0 \\
4 	& 2.0 		& 3.0 \\
5	& 3.0		& 4.0 \\
6	& 4.0		& 6.0 \\
7	& 6.0		& 9.0 \\
\hline
\end{tabular}
\end{center}
\label{tab:annuli}
\end{table}

\subsubsection{Data extraction and fitting}

\begin{figure*}
\begin{center}
\includegraphics[width=0.6\textwidth]{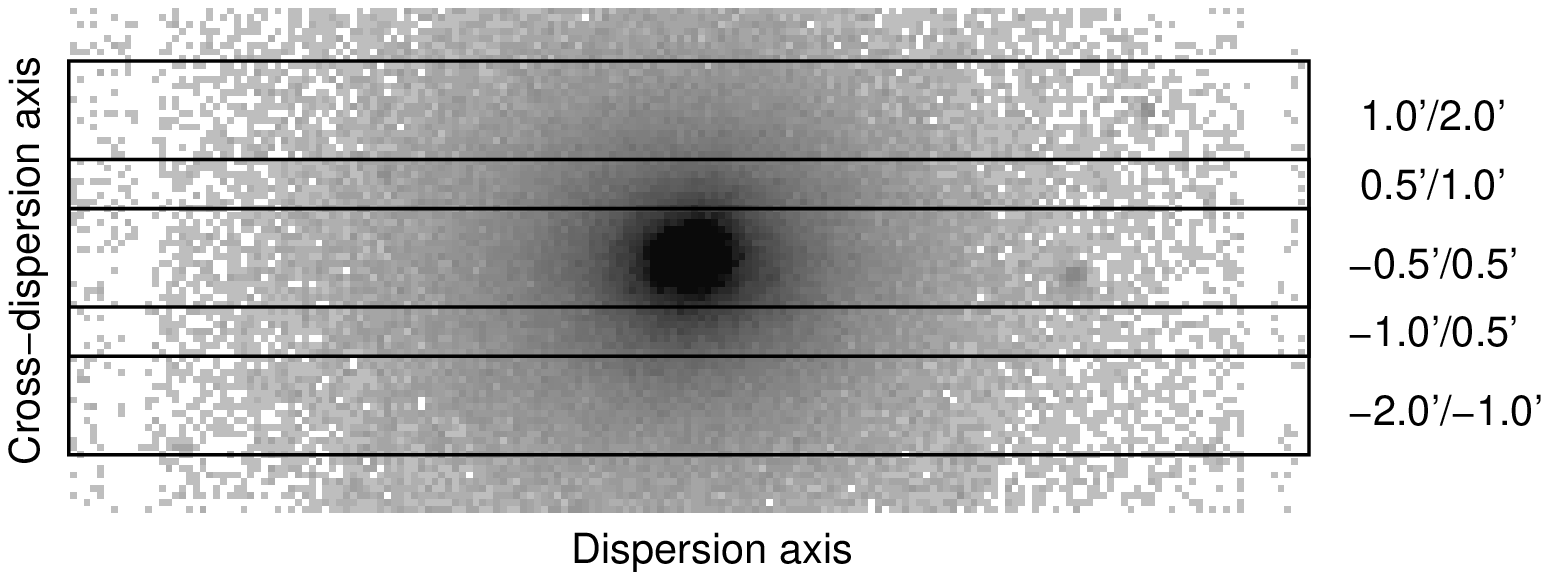}
\includegraphics[width=\textwidth]{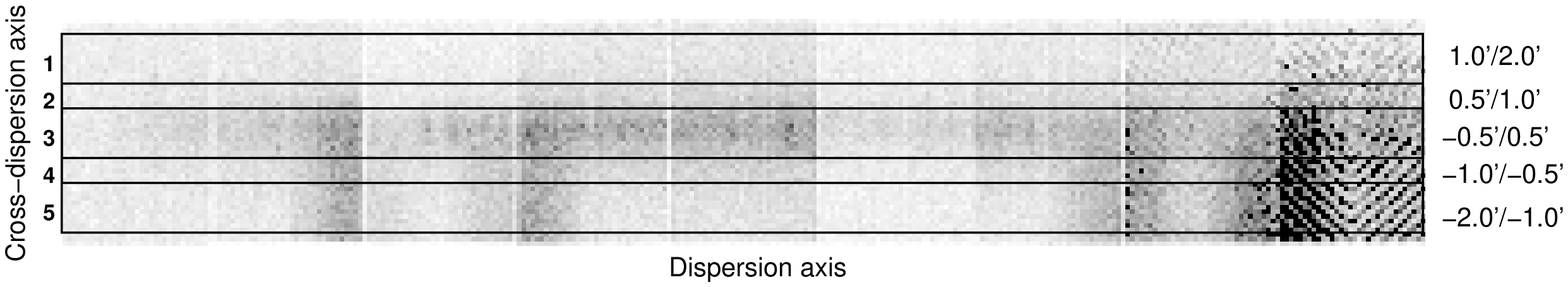}
\end{center}
\caption{RGS extraction regions projected on top of a MOS1 image ({\it top panel})
and projected on top of a combined RGS1/RGS2 detector image ({\it bottom panel}).
In the eighth and ninth CCD of RGS the instrumental effect of fixed-pattern noise
causes the fish-bone shaped patterns.}
\label{fig:rgsregions}
\end{figure*}

As we are interested in the spectral parameters as a function of radius, we extract spectra from annuli
as defined in Table~\ref{tab:annuli}.
To gain statistics we choose a width for the annuli which is $\ge$ 30$\arcsec$.
By using these relatively wide annuli, we are also less sensitive to the energy dependent shape of the 
PSF and therefore we neglect this effect in the rest of the analysis. Because of small calibration 
differences between MOS and pn we include a 5\% systematic error on the source and background spectra. 
This error is added in quadrature to the poissonian error of the data points.

\subsection{RGS analysis}
\label{sec:rgs_analysis}

We extract the RGS spectra following the method described in \citet{tamura2001b}.
In addition, we do spatially resolved spectroscopy using the RGS data:
with RGS it is possible to extract spectra from different regions in the inner
4$\arcmin$ of the cluster. The different regions can only be separated if they 
lie along the cross-dispersion axis of the instrument. We select the events from 
several rectangular areas on the CCD strip in the cross-dispersion direction. 
In Fig.~\ref{fig:rgsregions} we show how these extraction regions are projected on 
the sky (MOS1, upper panel) and on the chips of the RGS (lower panel).

Because the cluster fills the entire field-of-view of the RGS, we need a blank field observation
to extract the background spectrum. For this observation, which was taken prior to the cooling of RGS1 but after
cooling of RGS2, we choose a Lockman Hole observation with an effective exposure time of 100 ks performed just 3 XMM-Newton 
orbits after our observation, also still before the cooling of RGS1. This way we minimise systematic effects due to
hot pixels and instrument response.
The flare subtraction is analogous to the method used with EPIC, but now we use the events from CCD 9 
outside the central area. We select only events that have a position greater than 30$\arcsec$ from the
dispersion axis to make the light curve.
This method was applied to both source and background datasets.

The RGS spectrometer operates without a slit. This means that all photons from within the (in our case) $5\arcmin$ 
$\times$ $\sim12\arcmin$ field of view end up in the final spectrum, but not necessarily at the right wavelength. 
Only photons which are emitted in the cluster centre end up at the dispersion coordinate which corresponds to the correct
wavelength. If the photon originates from the outskirts at an angle $\theta$ (projected on the
dispersion axis) from the cluster centre, then the instrument will register it at a different dispersion coordinate
and assign a wavelength to it which is shifted with respect to the true wavelength. This shift in wavelength depends 
linearly on incidence angle $\theta$ (arcmin) projected on the dispersion axis: 
\begin{equation}
\Delta\lambda = 0.138\AA ~ \Delta\theta.
\label{eq:rgs}
\end{equation}
Because of this effect, the line-emission appears to be broadened depending on the spatial extent of the source 
along the dispersion direction \citep[see][~for a complete discussion about grating responses]{davis2001}.

In order to describe the data properly, the spectral fits need to account for the effect described above. 
In practice, this is accomplished by convolving the spectral models with the surface brightness profile of the 
source along the dispersion direction \citep{tamura2004}. Therefore, we extract the cluster intensity profile 
from MOS1 along the dispersion direction of RGS.
For each extraction region we can convert the spatial profile, which is a function of $\theta$, into a line
profile in \AA ~using Eq.~\ref{eq:rgs}. We convolve this profile with the model spectrum during spectral fitting.
However, the width of the measured line profile can be different from the line profile that we derive from MOS1,
because the surface brightness at the wavelength of a spectral line is not necessarily the same as the surface 
brightness in a broad  energy band. 
To take this effect into account, we multiply the wavelength axis of the derived line profile with a scale factor. 
A scale factor of 1.0 corresponds directly to the wavelength scale of the profile 
we derive from the continuum. A scale factor of 2.0 stretches the line profile in wavelength space 
by a factor of two compared to the original line profile. In this way, we are able to 
fit the widths of the line profiles and to have a measure of the spatial extent of the line-emission region.

\section{Spectral Models}
\label{sec:specmodel}

In our analysis we fit several models to the spectra using the SPEX package \citep{kaastra1996}. These models can be 
a combination of a number of thermal models (MEKAL) and a power-law model. Two models,
however, include a more sophisticated combination of thermal models that we call 
Differential Emission Measure (DEM) models. Previous papers \citep[e.g.][]{peterson2003,kaastra2004,deplaa2004}
show that many clusters can be fitted better when we use a distribution of temperatures
instead of a single temperature model. Actually, we expect that the plasma
within our annuli should contain multiple temperatures and not one. An observational proof of this
is given in \citet{werner2005}. In the case of S\'ersic 159-03 we fit two types 
of distributions: a truncated power-law distribution ({\it wdem}) and a 
Gaussian distribution ({\it gdem}).

\subsection{WDEM}

We use the so-called {\it wdem} model, where the
emission measure, $Y = \int n_e n_H dV$, of a number of thermal components is
distributed as a truncated power law. This is shown in Eq.~(\ref{eq:dy_dt})
adapted from \citet{kaastra2004}:
\begin{equation}
\frac{dY}{dT} = \left\{ \begin{array}{ll}
cT^{1/\alpha} & \hspace{1.0cm} \beta T_{\mathrm{max}} \le T < T_{\mathrm{max}} \\
0 & \hspace{1.0cm} T > T_{\mathrm{max}} \lor T < \beta T_{\mathrm{max}} .\\
\end{array} \right.
\label{eq:dy_dt}
\end{equation}
This distribution is cut off at a fraction of $T_{\mathrm{max}}$ which is $\beta T_{\mathrm{max}}$.
The value of $\beta$ is set to 0.1 in this study. The model above is an empirical parametrisation
of the DEM distribution found in the core of many clusters. In this form the limit
$\alpha \to 0$ yields the isothermal model at $T_{\mathrm{max}}$.

In order to compare the outcome of the {\it wdem} model with single-temperature models, we can 
calculate the emission weighted mean of the DEM distribution. The mean temperature 
$kT_{\mathrm{mean}}$ follows from Eq.~\ref{eq:em_weight}:
\begin{equation}
T_{\mathrm{mean}} = \frac{\int T \frac{dY}{dT} dT}{\int \frac{dY}{dT} dT}.
\label{eq:em_weight}
\end{equation}
When we integrate this equation between $\beta T_{\mathrm{max}}$ and $T_{\mathrm{max}}$, 
we obtain a direct relation between $T_{\mathrm{mean}}$ and 
$T_{\mathrm{max}}$ as a function of $\alpha$ and $\beta$:
\begin{equation}
T_{\mathrm{mean}} = \frac{(1 + 1/\alpha)}{(2 + 1/\alpha)} \frac{(1 - \beta^{1/\alpha + 2})}{(1 - \beta^{1/\alpha + 1})} T_{\mathrm{max}}.
\label{eq:wdem_mean}
\end{equation}
The values for $kT_{\mathrm{mean}}$ we present in this paper are calculated using Eq.~\ref{eq:wdem_mean}.

A detailed comparison of the {\it wdem} model with the classical cooling-flow model can be found in
\citet{deplaa2005a}. In general the {\it wdem} model contains less cool gas than the classical
cooling-flow model, which is consistent with recent observations \citep{peterson2001,peterson2003}.

\subsection{GDEM}

Another DEM model that we use is a Gaussian differential emission measure distribution, {\it gdem}, 
in $\mathrm{log}~T$:
\begin{equation}
Y(x) = \frac{Y_0}{\sigma_{\mathrm{T}} \sqrt{2\pi }} \mathrm{e}^{-(x-x_0)^2 / 2\sigma^2_{\mathrm{T}}}.
\end{equation} 
In this equation $x=\mathrm{log}~T$ and $x_0=\mathrm{log}~T_0$ where $T_0$ is the average temperature 
of the distribution. The width of the Gaussian is $\sigma_{\mathrm{T}}$. Compared to the {\it wdem} 
model this distribution contains more emission measure at higher temperatures.

\section{Results}

\begin{figure*}[t]
\includegraphics[width=\textwidth]{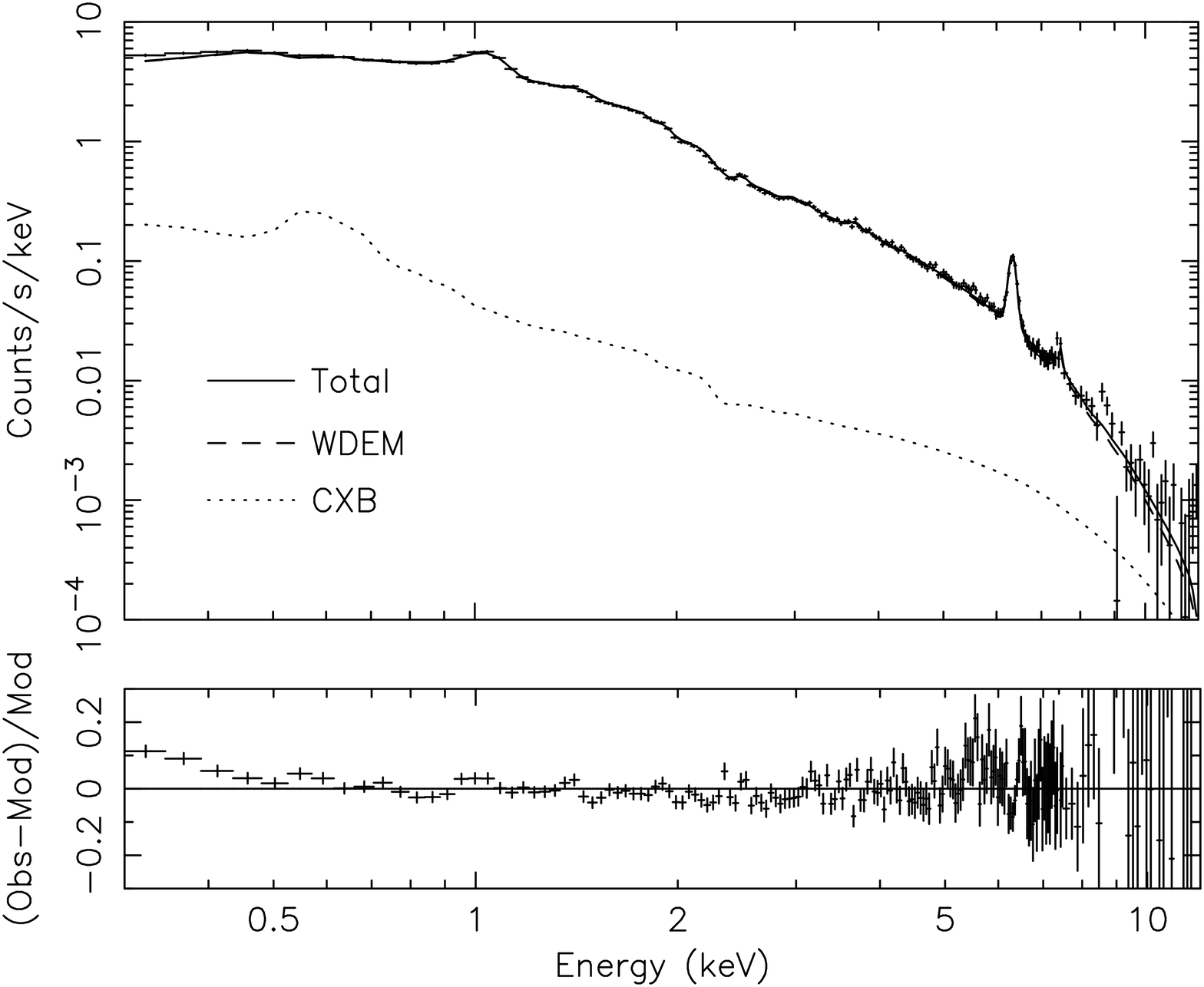}
\caption{
EPIC spectrum extracted from a circular region of 4$\arcmin$ around the core. The total best-fit 
model, the {\it wdem} model component and the background component are shown.}
\label{fig:epic_total}
\end{figure*}

\subsection{EPIC/RGS spectral fitting}
\label{sec:epicrgs}

In this section we fit both the EPIC and RGS spectra with the models described in 
Sect.~\ref{sec:specmodel}. From the fits we obtain temperatures and abundances for which
we also derive spatial information.  
In all the fits we fix $N_{\mathrm{H}}$ to 1.79 $\times$ 10$^{20}$ cm$^{-2}$ which is the 
value deduced from \ion{H}{i} data \citep{dickey1990}. This value is the same as the one used in 
\citet{kaastra2004}. We use the \citet{verner1996} 
cross-sections in our absorption model. Throughout the paper we use errors at the $\Delta\chi^2 =1$
(68\%) level for one interesting parameter. 

The EPIC spectra are rebinned to the 
optimal binning of $\sim$1/3 FWHM of the resolution of the instrument.
We use the response matrix to derive the FWHM for every energy. The spectra are fitted over the 0.3--10 keV range.
We are aware that below 0.6 keV there are some calibration uncertainties of the order of 
5--10\% when using SAS 6.1.0. This is partially accounted for by the 5\% systematic error that we
add to the spectra during fitting. The remaining calibration uncertainties are still small
compared to the uncertainties in the soft X-ray background. Because of the importance of the 
soft-excess problem and the role of O VII we choose to include the 0.3--0.6 keV energy band 
in our fits.

For RGS we discuss the spectrum which was extracted from a 4$\arcmin$ wide strip in the cross-dispersion
direction, with which we obtain the highest signal-to-noise. In addition we also present 
spatially-resolved spectra extracted from smaller strips in the cross-dispersion direction. 
This provides a radial profile of the temperature and abundances in the core region (0--2$\arcmin$).    

After preliminary analysis, the RGS spectra show a discrepancy with any thermal model 
around 29--33 \AA. The data points are significantly below the models in this interval, 
as can be seen in Fig.~\ref{fig:rgs_wdem}. Since the wavelength range coincides with 
a read-out node of CCD2, we believe this feature is instrumental in nature. If we 
perform the same analysis on an earlier observation of S\'ersic 159-03 the discrepancy disappears,
which supports our view. In \citet{peterson2003} some other RGS observations of clusters 
of galaxies show the same problem, namely \object{Hydra A}, \object{Abell 496} and 
\object{MKW 3s}. A large study of many RGS observations shows that about $\sim$ 5\%
of the observations show an anomaly in the count rate of CCD2 (A. Pollock, priv. comm.). 
We therefore believe the feature we observe is instrumental in nature and we ignore the 29--33 \AA~ 
interval in our analysis.        

\begin{table}[tb]
\caption{Fit results for an EPIC spectrum extracted from a circle with a radius of 4$\arcmin$ and centred on the core.
Fluxes are calculated over the 0.3--10 keV range and presented in 10$^{-10}$ erg cm$^{-2}$ s$^{-1}$ deg$^{-2}$.
Emission measure ($Y_{\mathrm{thermal}} = \int n_e n_H \mathrm{d}V$)
is given in 10$^{66}$ cm$^{-3}$ and $Y_{\mathrm{pow}}$ is given in 10$^{51}$ ph s$^{-1}$ keV$^{-1}$ at 1 keV.}
\begin{center}
\begin{tabular}{lccc}
\hline\hline
Parameter	& single-temp 		& {\it wdem}-model	& {\it gdem}-model 	\\	
\hline
$Y_{\mathrm{thermal}}$
		& 20.31 $\pm$ 0.15	& 20.55 $\pm$ 0.15	& 21.07 $\pm$ 0.15	\\
$F_{\mathrm{thermal}}$
		& 19.26 $\pm$ 0.14 	& 19.40 $\pm$ 0.14	& 19.46 $\pm$ 0.15	\\
$kT$		& 2.568 $\pm$ 0.009 	&			& 2.472 $\pm$ 0.010	\\
$kT_{\mathrm{mean}}$&			& 2.60 $\pm$ 0.03	&			\\
$kT_{\mathrm{max}}$& 			& 3.41 $\pm$ 0.03	& 			\\
$\alpha$	&			& 0.45 $\pm$ 0.02	&			\\
$\sigma_{T}$	&			&			& 0.226 $\pm$ 0.005	\\
O		& 0.36 $\pm$ 0.05	& 0.30 $\pm$ 0.04	& 0.19 $\pm$ 0.03	\\
Ne		& 0.89 $\pm$ 0.08	& 0.16 $\pm$ 0.08	& 0.000 $\pm$ 0.013	\\
Mg		& 0.02$\pm$0.04		& 0.11 $\pm$ 0.04 	& 0.08 $\pm$ 0.04	\\
Si		& 0.196 $\pm$ 0.017	& 0.239 $\pm$ 0.017	& 0.252 $\pm$ 0.018	\\
S		& 0.122 $\pm$ 0.017	& 0.18 $\pm$ 0.02	& 0.20 $\pm$ 0.02	\\
Ar		& 			& 0.14 $\pm$ 0.04	& 0.19 $\pm$ 0.05	\\
Ca 		& 0.31 $\pm$ 0.05	& 0.36 $\pm$ 0.05	& 0.44 $\pm$ 0.06	\\
Fe		& 0.360 $\pm$ 0.007	& 0.346 $\pm$ 0.008 	& 0.242 $\pm$ 0.005	\\
Ni		& 0.42 $\pm$ 0.09	& 0.45 $\pm$ 0.08	& 0.35 $\pm$ 0.08	\\
\hline
$\chi^2$ / dof	& 1228 / 916		& 1017 / 915		& 949 / 915		\\	
\hline
\end{tabular}
\label{tab:4arcmin}
\end{center}
\end{table}

\begin{table}[tb]
\caption{Fit results for RGS spectra extracted from a 4$\arcmin$ wide strip and fitted over
a 8--38 \AA~range ignoring data from CCD2 (see text). 
Emission measures ($Y = \int n_e n_H \mathrm{d}V$) are given in 10$^{66}$ cm$^{-3}$.
The iron abundance is given with respect to solar abundances.
The scale factors for oxygen and iron mentioned 
below are those explained in Sect.~\ref{sec:rgs_analysis}.}
\begin{center}
\begin{tabular}{l|ccc}
\hline\hline
Parameter 	& single-temp	 	& {\it wdem}-model 	& {\it gdem}-model \\
\hline
$Y$		& 7.27$\pm$0.06		& 7.08$\pm$0.07		& 7.25$\pm$0.06	\\
$kT$ (keV)	& 2.6$\pm$0.09		&  			& 3.25$\pm$0.18	\\
$kT_{\mathrm{mean}}$ (keV) &		& 3.04 $\pm$ 0.16	&		\\
$kT_{\mathrm{max}}$ (keV) &		& 4.0$\pm$0.2		&		\\
$\alpha$	& 			& 0.46$\pm$0.03		& 	\\
$\sigma$	&			& 			& 0.28$\pm$0.02	\\
C/Fe		& 0.00$\pm$0.19		& 0.0$\pm$0.4		& 0.0$\pm$0.3		\\
N/Fe		& 0.00$\pm$0.18		& 0.1$\pm$0.5		& 0.0$\pm$0.5		\\
O/Fe 		& 0.85$\pm$0.10		& 0.87$\pm$0.10		& 0.91$\pm$0.13	\\
Ne/Fe		& 1.04$\pm$0.15		& 0.73$\pm$0.12		& 0.71$\pm$0.14	\\
Mg/Fe		& 0.32$\pm$0.11		& 0.35$\pm$0.11		& 0.32$\pm$0.12	\\
Fe		& 0.98$\pm$0.06		& 1.20$\pm$0.08		& 1.14$\pm$0.09 	\\
Scale (O)	& 1.1$\pm$0.2		& 1.6$\pm$0.3		& 1.5$\pm$0.3	\\
Scale (Fe)	& 0.35$\pm$0.05		& 0.36$\pm$0.05		& 0.31$\pm$0.05	\\
\hline
$\chi^2$ / dof	& 997 / 856		& 925 / 854		& 963 / 855		\\
\hline
\end{tabular}
\label{tab:rgs_240}
\end{center}
\end{table}

\subsubsection{Integrated spectrum within 4$\arcmin$}
\label{sec:intspec}

In Table~\ref{tab:4arcmin} we show the results from fits to a spectrum extracted from a 
circular region which has a radius of 4$\arcmin$ and is centred on the core. In this circular region
the cluster signal is well above the background. We can use the high signal-to-noise ratio of this spectrum 
to obtain accurate values for all the fitted parameters. The $\chi^2$ value of the single-temperature 
fit (1228/916) indicates that a single-temperature model is potentially not the best description of the data.
Multi-temperature models, however, do produce a $\chi_r^2$ close to 1.0. Hence, our abundance 
analysis will be based on the multi-temperature models. There are other models that fit
the data equally well. These alternative models will be discussed in Sect.~\ref{sec:hardsoft}. 
The potential presence of a relatively small contribution of other emission components however is of no 
consequence for the observed trends we present in this section.

\begin{figure*}[t]
\begin{minipage}{0.49\textwidth}
\includegraphics[width=\columnwidth]{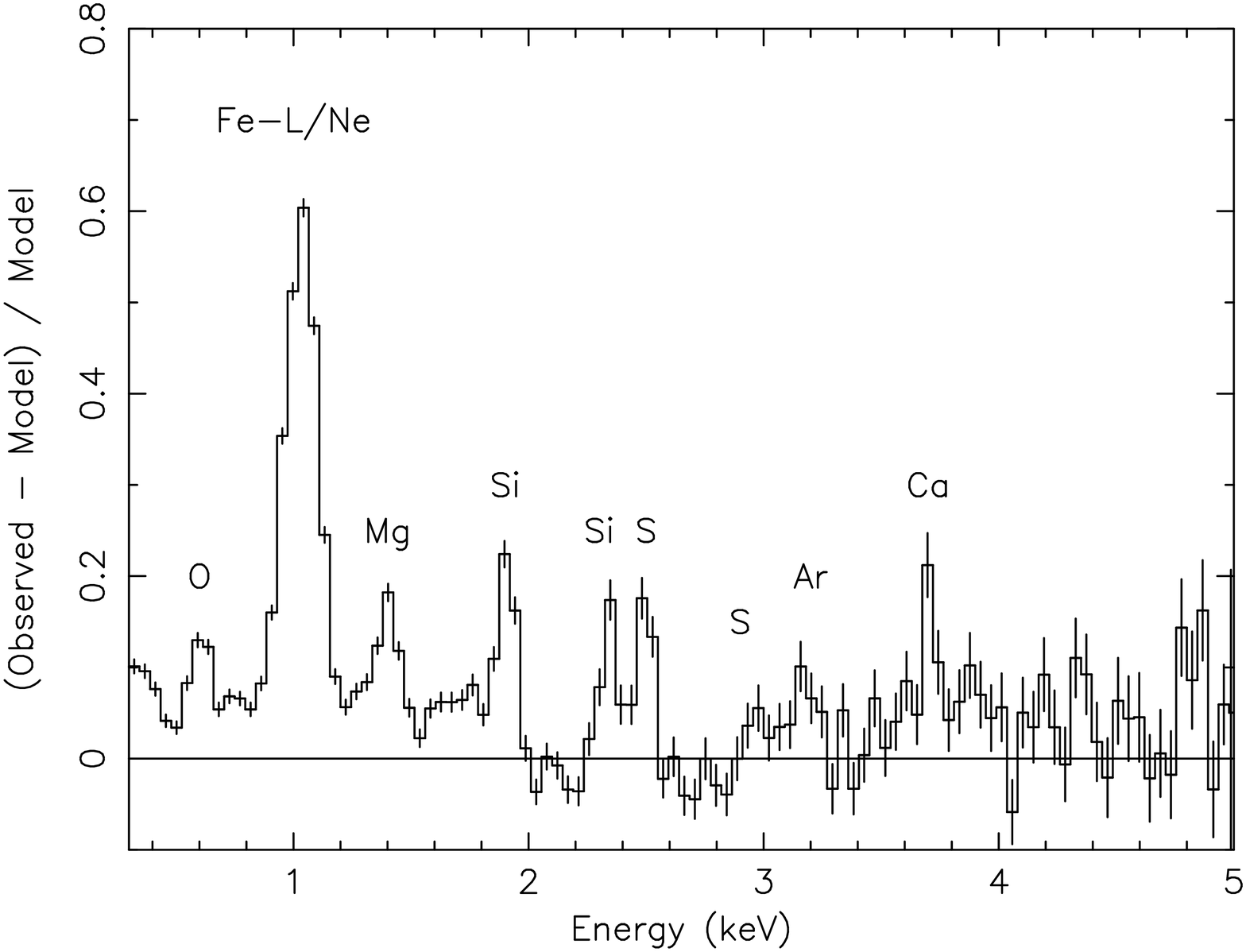}
\end{minipage}
\hspace{2mm}
\begin{minipage}{0.49\textwidth}
\includegraphics[width=\columnwidth]{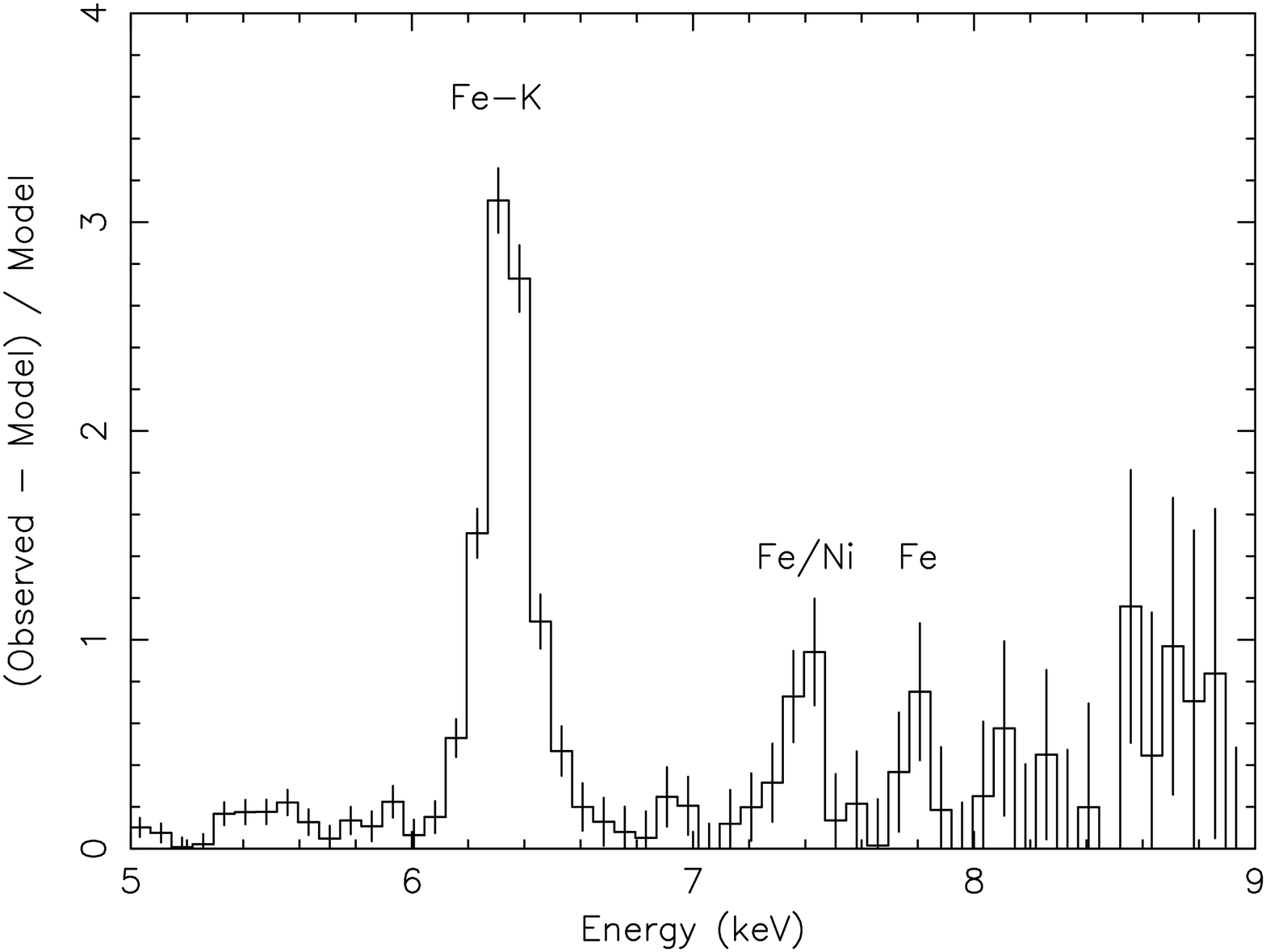}
\end{minipage}
\caption{Residuals of a {\it wdem} fit to the EPIC total spectrum extracted from a circle with a radius of 4$\arcmin$.
The line emission in the model is set to zero to show the line emission on top of the continuum. The left panel
shows the residuals from 0.3--5 keV and the right panel the 5--10 keV range.} 
\label{fig:epic_lines}
\end{figure*}

An example of a spectrum fitted with a DEM model is shown in Fig.~\ref{fig:epic_total}. 
When we set the line emission to zero in our model and plot the residuals, we obtain the plot shown 
in Fig.~\ref{fig:epic_lines}. The lines from all elements for which we fit abundances are evident in 
the spectrum. Especially the iron lines are very strong, but also silicon, sulfur and calcium have a high 
signal-to-noise. Oxygen, magnesium, argon and nickel are clearly visible: this plot shows 
the total spectrum; the strengths of these lines are weaker in individual annuli.

\begin{figure*}[t]
\includegraphics[width=\textwidth]{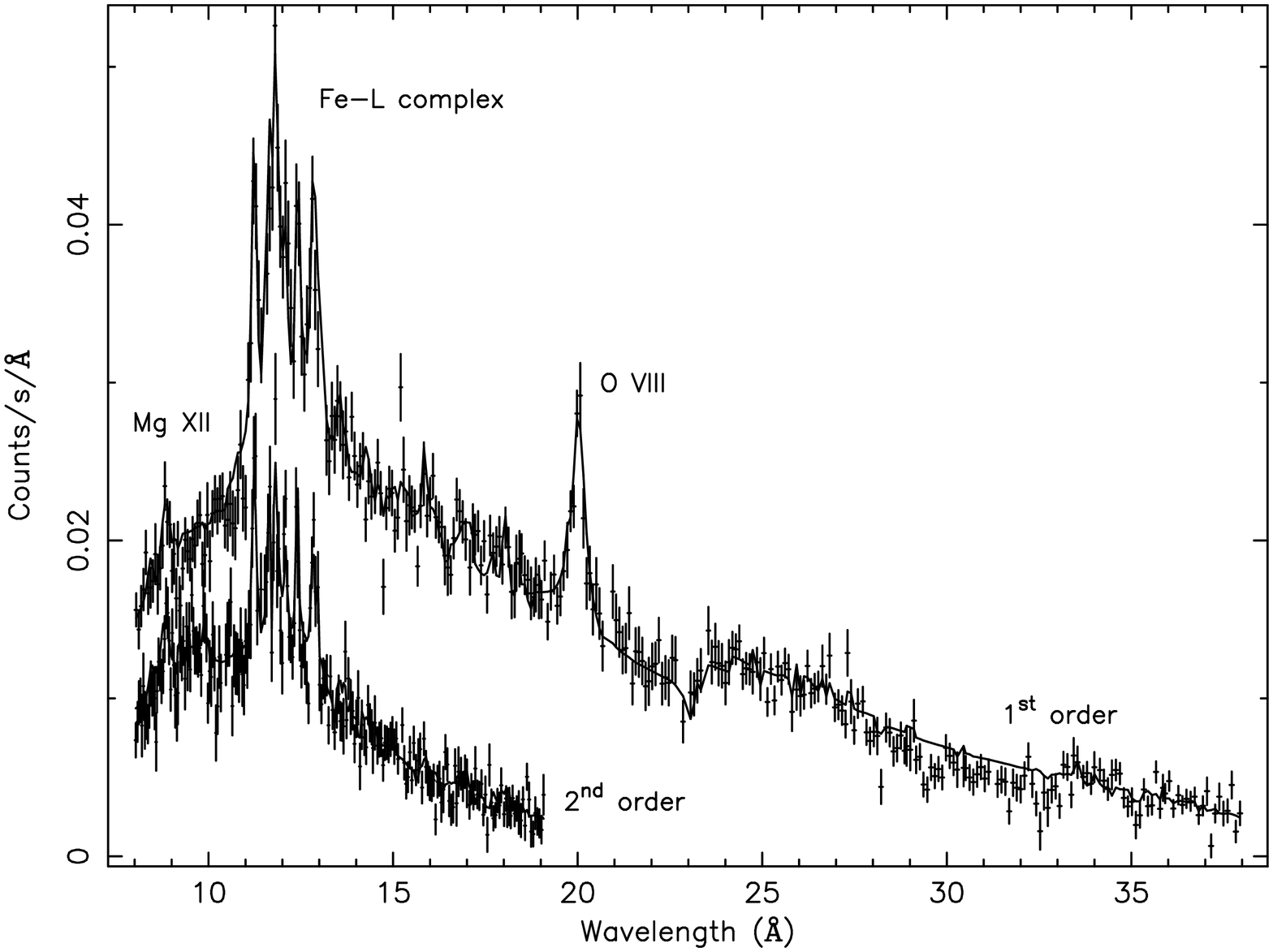}
\caption{~1$^{st}$ and 2$^{nd}$ order RGS spectrum of S\'ersic 159-03 extracted from a 4$\arcmin$ wide strip centred on the core. 
The continuous line represents the fitted {\it wdem} model. On the x-axis we show the observed wavelength.}
\label{fig:rgs_wdem}
\end{figure*}

The spectral models constrain the abundances very well, except for neon and magnesium, because these lines 
are blended with iron at this spectral resolution (see Fig.~\ref{fig:epic_lines}).
The determination of the oxygen abundance becomes difficult in the outer parts of the cluster.
Here, the oxygen in the galactic foreground emission starts to play a more important role and could bias
the abundance measurement. In the central region ($\lesssim$ 4$\arcmin$) the cluster flux is high enough
to get an accurate measurement. 
In general, all abundances are between 0.1 and 0.6 times their solar value (Table~\ref{tab:4arcmin}).
Only neon shows high values up to 1, but since the lines of this element are blended with the Fe-L 
complex, the value is highly correlated with the temperature distribution.

In Fig.~\ref{fig:rgs_wdem} we present the total RGS spectrum of S\'ersic 159-03 extracted from 
a 4$\arcmin$ wide strip in the cross-dispersion direction and centred on the core. Because
of the high statistics, we can also use the second order spectrum. In both orders the 
Fe-L spectral line complex between 10--15 \AA~and the Mg line near 9 \AA~are well resolved. Above
19 \AA~the first order spectrum shows a prominent \ion{O}{viii} Ly $\alpha$ line.

\begin{figure}
\includegraphics[width=1.0\columnwidth]{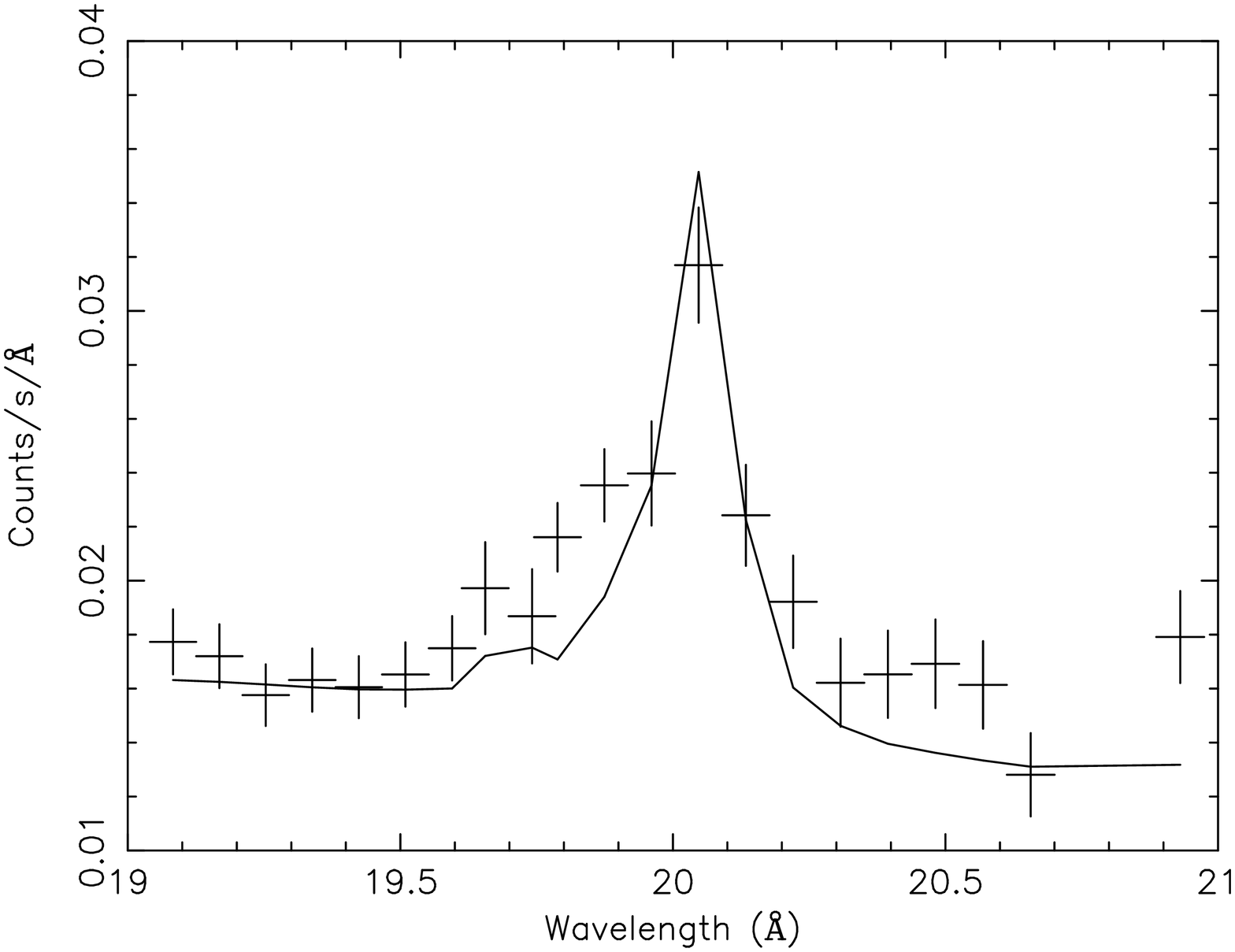}
\caption{RGS spectrum in the 19--21\AA~band featuring the \ion{O}{viii} Ly-$\alpha$ line.
The data points represent the measured spectrum, while the model line shows the line when 
the scale factor is frozen to $\sim$0.3, which is the best-fit value for the iron lines.}
\label{fig:rgs_width}
\end{figure}

The fit results for RGS are shown in Table~\ref{tab:rgs_240}. 
The elemental abundances determined from the different models are consistent
with each other within 3$\sigma$. Unfortunately, the carbon and nitrogen 
abundances are not well constrained.   
The width of the oxygen line is consistent with the width of the line profile derived from 
the continuum emission in MOS1, and the width of the iron lines is about 1/3 smaller (Fig.~\ref{fig:rgs_width}).
This suggests that the oxygen distribution is more extended across the cluster centre, 
while the iron abundance is strongly peaked in the centre. Looking at the $\chi^2$ values we see that 
the fit to the RGS data does not strongly prefer a DEM distribution over a single-temperature model.
All models are acceptable, contrary to the merging cluster \object{2A 0335+096} 
described by \citet{werner2005}.

\subsubsection{Radial profiles using thermal DEM models}

\begin{figure*}[!tb]
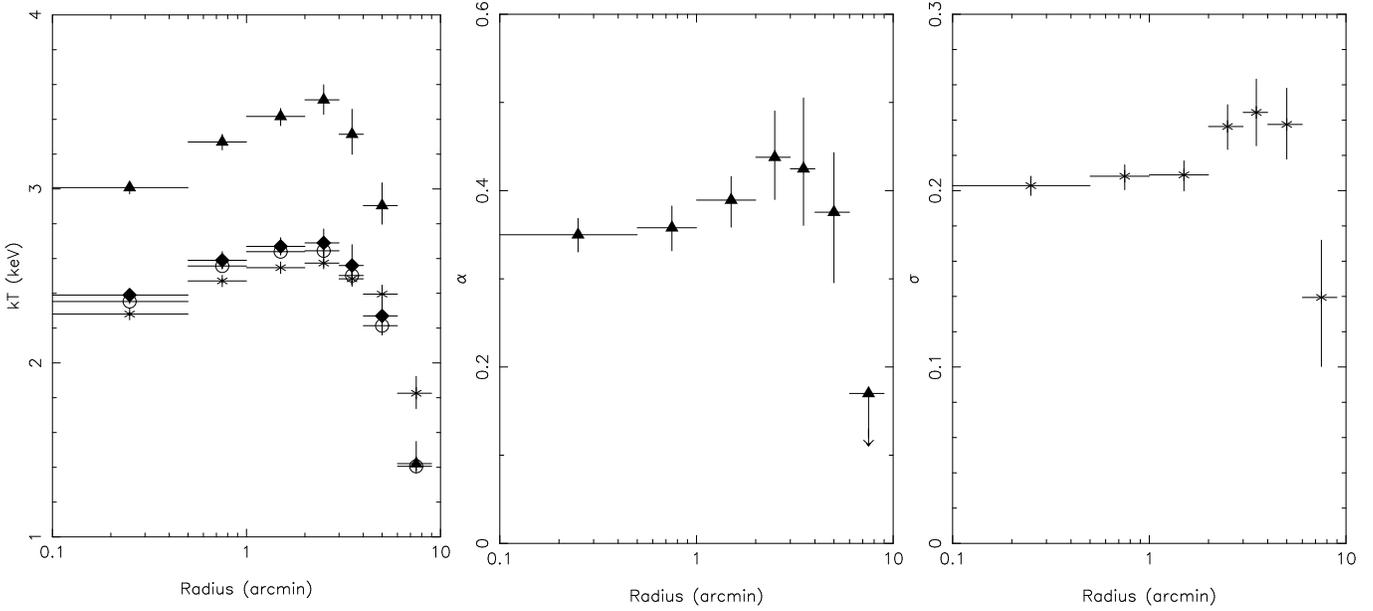

\begin{minipage}{0.33\textwidth}
\includegraphics[width=1.0\textwidth]{figure7a.ps}
\end{minipage}
\begin{minipage}{0.33\textwidth}
\includegraphics[width=1.0\textwidth]{figure7b.ps}
\end{minipage}
\begin{minipage}{0.33\textwidth}
\includegraphics[width=1.0\textwidth]{figure7c.ps}
\end{minipage}\\
\caption{DEM model fit results for the three models: single-temperature ($\bigcirc$), 
{\it wdem} ($\blacktriangle$) and {\it gdem} ($\ast$).  
In the left panel we show the radial temperature profiles for all models, including the 
$kT_{\mathrm{max}}$ ($\blacktriangle$) and $kT_{\mathrm{mean}}$ ($\blacklozenge$) from the 
{\it wdem} parameters. In the middle and right panel we show the 
radial profiles of the DEM parameters $\alpha$ and $\sigma_{\mathrm{T}}$.}
\label{fig:epic_temp}
\end{figure*}

By fitting the spectra extracted from the annuli for EPIC and from strips for RGS,
we can make radial profiles of temperatures and abundances. Again, we fit the spectra with a
single-temperature model, a {\it wdem} model and a {\it gdem} model
(Table~\ref{tab:epic}). In Fig.~\ref{fig:epic_temp} (left panel) we present the temperature profile 
obtained for these models. For {\it wdem} the maximum temperature ($kT_{\mathrm{max}}$) and mean temperature
($kT_{\mathrm{mean}}$) are shown. 
The profile shows a slight increase of the temperature within a 2$\arcmin$ radius. Beyond
3$\arcmin$ the temperature rapidly drops to about 1.5 keV. The single-temperature profile
is consistent with the profile derived from the earlier XMM-Newton observation by \citet{kaastra2001}.

The DEM parameters $\alpha$ ({\it wdem}, Fig.~\ref{fig:epic_temp} middle panel) and $\sigma_{\mathrm{T}}$ 
({\it gdem}, Fig.~\ref{fig:epic_temp} right panel) are quite constant, but
increase slightly in the outer parts, where the temperature gradient is large. This means that the 
fit needs a broader range of temperatures to fit the spectrum.

\begin{figure*}[!tb]
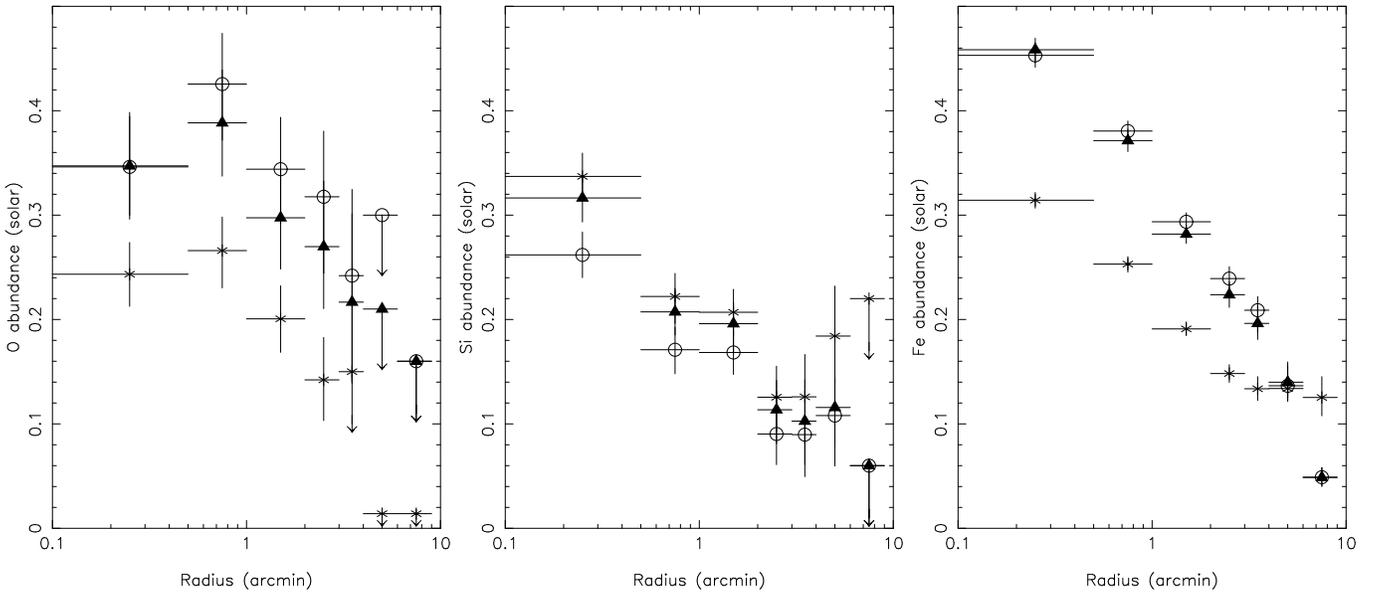

\begin{minipage}{0.33\textwidth}
\includegraphics[width=1.0\textwidth]{figure8a.ps}
\end{minipage}
\begin{minipage}{0.33\textwidth}
\includegraphics[width=1.0\textwidth]{figure8b.ps}
\end{minipage}
\begin{minipage}{0.33\textwidth}
\includegraphics[width=1.0\textwidth]{figure8c.ps}
\end{minipage}\\
\caption{Abundance profiles of oxygen, silicon and iron derived using three models: 
single-temperature ($\bigcirc$), {\it wdem} ($\blacktriangle$) and {\it gdem} ($\ast$).}
\label{fig:epic_abun}
\end{figure*}

From the fits we obtain radial profiles of the abundances of several elements. The radial profiles
of oxygen, silicon, and iron are shown in Fig.~\ref{fig:epic_abun}. 
We detect a clear decrease of the silicon and iron abundance with radius. 
For oxygen the situation is less clear: the O/Fe ratio in the 0--0.5$\arcmin$ bin is 
lower than the ratio in the 0.5--4$\arcmin$ interval at 3$\sigma$ confidence level. 
There is a hint of a decrease of the oxygen abundance like we see for iron, but a 
flat distribution cannot be excluded.

While silicon is quite well constrained independent of the DEM model used, 
the iron abundance derived from the {\it gdem} model is significantly different from the other two models. 
The absolute values of the {\it gdem} oxygen and iron abundances are consistent with the ones from silicon. 
The iron abundance from the single-temperature and {\it wdem}, however, show a steeper gradient. From these 
plots it is clear that the used temperature distribution is very important when measuring elemental abundances.

\setlength{\tabcolsep}{1.5mm}

\begin{table*}[!tb]
\caption{Fit results for spatially resolved EPIC spectra. The fitted models are
single-temperature CIE (s), {\it wdem} (w) and a Gaussian DEM model (g). 
Fluxes are calculated over the 0.3--10 keV range and are given in unit
erg s$^{-1}$ cm$^{-2}$ deg$^{-2}$. Emission measures ($Y = \int n_e n_H \mathrm{d}V$)
are given in 10$^{66}$ cm$^{-3}$.}
\begin{center}
\begin{scriptsize}
\begin{tabular}{l|c|ccccccc}
\hline\hline
& M		
& 0--0.5$\arcmin$	
& 0.5--1.0$\arcmin$	
& 1.0--2.0$\arcmin$ 
& 2.0--3.0$\arcmin$
& 3.0--4.0$\arcmin$
& 4.0--6.0$\arcmin$
& 6.0--9.0$\arcmin$
\\
\hline
$Y$	&
s 
 & 4.75$\pm$0.05
 & 5.55$\pm$0.05
 & 6.17$\pm$0.05
 & 2.88$\pm$0.03
 & 1.40$\pm$0.02
 & 1.62$\pm$0.04
 & 1.21$\pm$0.04
\\
 &  w
 & 4.72$\pm$0.05
 & 5.58$\pm$0.05
 & 6.23$\pm$0.05
 & 2.91$\pm$0.03
 & 1.42$\pm$0.02
 & 1.62$\pm$0.03
 & 1.21$\pm$0.03
\\
 &  g
 & 4.86$\pm$0.04
 & 5.72$\pm$0.05 
 & 6.36$\pm$0.05
 & 2.97$\pm$0.03
 & 1.44$\pm$0.02
 & 1.55$\pm$0.03
 & 0.94$\pm$0.04
\\
\hline
$F$	&
s 
 & (2.59$\pm$0.02)$\times$10$^{-8}$
 & (1.038$\pm$0.009)$\times$10$^{-8}$
 & (2.76$\pm$0.02)$\times$10$^{-9}$
 & (7.72$\pm$0.09)$\times$10$^{-10}$
 & (3.17$\pm$0.05)$\times$10$^{-10}$
 & (1.08$\pm$0.02)$\times$10$^{-10}$
 & (2.59$\pm$0.08)$\times$10$^{-11}$
\\
 & w
 & (2.60$\pm$0.03)$\times$10$^{-8}$
 & (1.042$\pm$0.010)$\times$10$^{-8}$
 & (2.77$\pm$0.02)$\times$10$^{-9}$   
 & (7.79$\pm$0.09)$\times$10$^{-10}$
 & (3.20$\pm$0.05)$\times$10$^{-10}$
 & (1.093$\pm$0.019)$\times$10$^{-10}$
 & (2.59$\pm$0.08)$\times$10$^{-11}$
\\
 & g
 & (2.61$\pm$0.02)$\times$10$^{-8}$
 & (1.047$\pm$0.010)$\times$10$^{-8}$
 & (2.78$\pm$0.02)$\times$10$^{-9}$  
 & (7.84$\pm$0.08)$\times$10$^{-10}$
 & (3.24$\pm$0.05)$\times$10$^{-10}$
 & (1.112$\pm$0.018)$\times$10$^{-10}$
 & (2.59$\pm$0.09)$\times$10$^{-11}$
\\
\hline
$kT$ 	&	
s 
 & 2.353$\pm$0.012
 & 2.557$\pm$0.013
 & 2.640$\pm$0.015 
 & 2.64$\pm$0.03 
 & 2.50$\pm$0.04
 & 2.21$\pm$0.05 
 & 1.41$\pm$0.04
\\
(keV) & g 
 & 2.280$\pm$0.011
 & 2.471$\pm$0.013 
 & 2.547$\pm$0.015
 & 2.57$\pm$0.02 
 & 2.48$\pm$0.04
 & 2.40$\pm$0.05
 & 1.83$\pm$0.09
\\
\hline
$kT_{\mathrm{mean}}$ 
& w		
 & 2.39$\pm$0.03
 & 2.59$\pm$0.05
 & 2.67$\pm$0.05
 & 2.69$\pm$0.08
 & 2.56$\pm$0.12
 & 2.27$\pm$0.11
 & 
\\
\hline
$kT_{\mathrm{max}}$ 
& w		
 & 3.01$\pm$0.03
 & 3.27$\pm$0.04
 & 3.42$\pm$0.05
 & 3.51$\pm$0.08
 & 3.31$\pm$0.13
 & 2.90$\pm$0.12
 & 1.42$\pm$0.08
\\
\hline
$\alpha$	
& w
 & 0.350$\pm$0.019
 & 0.36$\pm$0.03
 & 0.39$\pm$0.03
 & 0.44$\pm$0.05
 & 0.42$\pm$0.07
 & 0.38$\pm$0.07
 & 0.01$\pm$0.09
\\
\hline
$\sigma_{\mathrm{T}}$	
& g		
 & 0.203$\pm$0.005
 & 0.208$\pm$0.007
 & 0.209$\pm$0.008
 & 0.236$\pm$0.012
 & 0.244$\pm$0.019
 & 0.24$\pm$0.02
 & 0.14$\pm$0.04
\\
\hline
O/Fe &
s &
0.76$\pm$0.12&
1.12$\pm$0.13&
1.17$\pm$0.17&
1.3$\pm$0.3 &
1.2$\pm$0.4 &
0.2$\pm$1.0 &
0.0$\pm$1.7 
\\
& w &
0.76$\pm$0.11&
1.05$\pm$0.14&
1.06$\pm$0.17&
1.2$\pm$0.3 &
1.1$\pm$0.4 &
0.4$\pm$0.6 &
0.0$\pm$1.7 
\\
& g &
0.77$\pm$0.10 &
1.05$\pm$0.13 &
1.05$\pm$0.17 &
1.0$\pm$0.3 &
0.4$\pm$0.4 &
0.00$\pm$0.05 &
0.00$\pm$0.06 
\\
\hline
Ne/Fe &
s &
2.6$\pm$0.2 &
2.0$\pm$0.2 &
1.4$\pm$0.3 &
1.4$\pm$0.4&
1.5$\pm$0.6&
1.7$\pm$1.1&
0.0$\pm$0.9
\\
& w &
0.9$\pm$0.2&
0.4$\pm$0.3&
0.00$\pm$0.15&
0.0$\pm$0.2&
0.0$\pm$0.5&
0.2$\pm$0.9&
0.0$\pm$0.9
\\
& g &
0.00$\pm$0.16&
0.00$\pm$0.07&
0.00$\pm$0.05&
0.00$\pm$0.09&
0.0 $\pm$0.2 &
0.0 $\pm$0.5 &
0.0 $\pm$0.7 
\\
\hline
Mg/Fe &
s &
0.10$\pm$0.10&
0.00$\pm$0.12&
0.00$\pm$0.08&
0.01$\pm$0.19&
0.00$\pm$0.09&
0.1$\pm$0.5 &
0.0$\pm$0.4 
\\
& w &
0.29$\pm$0.10&
0.18$\pm$0.13&
0.13$\pm$0.15&
0.3$\pm$0.3 &
0.00$\pm$0.11 &
0.3$\pm$0.5 &
0.0$\pm$0.4 
\\
& g &
0.34$\pm$0.14&
0.20$\pm$0.18&
0.1$\pm$0.2&
0.5$\pm$0.4&
0.0$\pm$0.2&
1.3$\pm$0.6&
0.0$\pm$0.7
\\
\hline
Si/Fe &
s &
0.58$\pm$0.05&
0.45$\pm$0.06&
0.57$\pm$0.07&
0.38$\pm$0.12&
0.43$\pm$0.09&
0.8$\pm$0.3 &
0.5$\pm$0.5 
\\
& w &
0.69$\pm$0.05&
0.56$\pm$0.06&
0.70$\pm$0.08&
0.51$\pm$0.13&
0.5$\pm$0.2&
0.8$\pm$0.2&
0.4$\pm$0.5
\\
& g &
1.07$\pm$0.08&
0.88$\pm$0.09&
1.08$\pm$0.12&
0.8$\pm$0.2 &
0.9$\pm$0.3 &
1.4$\pm$0.4 &
0.7$\pm$0.6 
\\
\hline
S/Fe &
s &
0.43$\pm$0.06&
0.34$\pm$0.07&
0.19$\pm$0.09&
0.22$\pm$0.16&
0.3$\pm$0.3 &
0.97$\pm$0.17 &
5.6$\pm$1.7 
\\
& w &
0.56$\pm$0.06&
0.47$\pm$0.08&
0.35$\pm$0.09&
0.40$\pm$0.15&
0.5$\pm$0.2 &
1.0$\pm$0.4 &
5.6$\pm$1.8 
\\
& g &
0.90$\pm$0.09&
0.76$\pm$0.11&
0.57$\pm$0.13&
0.7$\pm$0.2 &
0.9$\pm$0.4 &
1.3$\pm$0.5 &
3.5$\pm$1.2 
\\
\hline
Ar/Fe &
s &
0.02$\pm$0.13&
0.20$\pm$0.17&
0.4$\pm$0.2 &
0.0$\pm$0.4 &
0.6$\pm$0.7&
1.9$\pm$1.1&
16.8$\pm$6.4 
\\
& w &
0.19$\pm$0.14&
0.39$\pm$0.18&
0.7$\pm$0.2&
0.2$\pm$0.4&
0.8$\pm$0.7&
1.8$\pm$1.1&
17.0$\pm$6.4 
\\
& g &
0.4$\pm$0.2&
0.7$\pm$0.3&
1.2$\pm$0.4&
0.4$\pm$0.7&
1.4$\pm$1.2&
1.6$\pm$1.5&
5.4$\pm$3.5
\\
\hline
Ca/Fe &
s &
0.59$\pm$0.18&
1.1$\pm$0.2&
1.1$\pm$0.3&
1.0$\pm$0.5&
1.6$\pm$1.0&
3.6$\pm$1.5&
36.6$\pm$15.6
\\
& w &
0.67$\pm$0.18&
1.3$\pm$0.2&
1.2$\pm$0.3&
1.0$\pm$0.6&
1.6$\pm$1.0&
3.0$\pm$1.6&
36.2$\pm$15.9 
\\
& g &
1.2$\pm$0.3&
2.3$\pm$0.4&
2.2$\pm$0.5&
1.7$\pm$1.1&
2.2$\pm$1.8&
0.4$\pm$2.3&
0.0$\pm$6.5
\\
\hline
Fe &
s &
0.453$\pm$0.012&
0.381$\pm$0.010&
0.294$\pm$0.009&
0.239$\pm$0.011&
0.209$\pm$0.013&
0.14$\pm$0.02 &
0.049$\pm$0.009 
\\
& w &
0.458$\pm$0.011&
0.371$\pm$0.011&
0.282$\pm$0.009&
0.224$\pm$0.012&
0.196$\pm$0.017&
0.140$\pm$0.016&
0.049$\pm$0.009
\\
& g &
0.314$\pm$0.007&
0.253$\pm$0.007&
0.191$\pm$0.007&
0.148$\pm$0.008&
0.134$\pm$0.012&
0.134$\pm$0.012&
0.13$\pm$0.02
\\
\hline
Ni/Fe &
s &
1.0$\pm$0.2&
0.9$\pm$0.2&
0.0$\pm$0.3&
0.0$\pm$0.2&
0.3$\pm$0.8&
0.0$\pm$0.3&
0.0$\pm$2.3
\\
& w &
1.2$\pm$0.2&
1.0$\pm$0.3&
0.2$\pm$0.3&
0.0$\pm$0.5&
1.1$\pm$1.0&
0.0$\pm$0.3&
0.0$\pm$2.4
\\
& g &
1.4$\pm$0.3&
1.1$\pm$0.4&
0.0$\pm$0.5&
0.0$\pm$0.8&
2.8$\pm$1.5&
0.3$\pm$1.6&
6.0$\pm$2.5
\\
\hline
$\chi^2$ / dof & s 
 & 1213 / 829
 & 1084 / 846
 & 1240 / 852
 & 1191 / 804
 & 1270 / 772
 & 1113 / 777
 & 1045 / 757
\\
 & w 
 & 1045 / 828 
 & 969 / 845 
 & 1136 / 851
 & 1109 / 803
 & 1218 / 771
 & 1095 / 776
 & 1045 / 756
\\
 & g 
 & 981 / 828
 & 890 / 845
 & 1083 / 851
 & 1034 / 803
 & 1131 / 771
 & 1042 / 776
 & 1079 / 756
\\
\hline
\end{tabular}
\end{scriptsize}
\label{tab:epic}
\end{center}
\end{table*}

\setlength{\tabcolsep}{1.5mm}

An overview of all the fitted parameters is given in Table~\ref{tab:epic}. 
Apart from the elements we discussed in the previous paragraph, only sulfur is reasonably well constrained
up to 9$\arcmin$. In the outermost spatial bin, however, the abundance increases
to unphysical values. In the core region sulfur shows the same profile as
silicon. The other elements are less well constrained. Neon is always difficult to measure
at CCD resolution because it is blended with the Fe-L complex near 1 keV. Calcium and nickel
are only constrained in the core, but the derived values are not affected by systematics
due to different DEM models. In every radial bin the values are consistent with each other.
The magnesium and argon abundances are poorly determined. Only in the core region some points 
are measured at a significance larger than 2$\sigma$. In general these two elements have upper 
limits of about 0.2 times the solar abundances.

The $\chi^2$ values for the three models are also shown in Table~\ref{tab:epic}. In the core 
region all fits are acceptable when we use the condition $\chi_r^2 < 1.5$.
However, the $\chi_r^2$ for the single temperature model in the core is much less than in 
\citet{werner2005} in the cluster 2A 0335+096. In 
the outer parts multi-temperature fitting still results in a better fit, but we do not see 
significant differences between the models.

From RGS we also extract rectangular regions in the cross-dispersion direction of the instrument. 
The regions we use are defined in Sect.~\ref{sec:rgs_analysis}.
Because the cluster does not show significant spatial asymmetries in the cross-dispersion direction, 
we add the spectra extracted from regions with equal distance to the dispersion axis: region 1 + 5 
and 2 + 4 as defined in the lower panel of Fig.~\ref{fig:rgsregions}.
This way we are able to derive radial profiles of the core up to a radius of 2$\arcmin$.
The best fit values for three models are presented in Table~\ref{tab:rgs_sp}. The 
fit results show signs of a temperature decrement in the core. However, the
temperatures we determine from RGS are systematically higher than those from EPIC.
This is not surprising, because the RGS spectrum also contains emission 
from the hot gas just outside 2$\arcmin$ which falls within the rectangular field-of-view.
From the {\it wdem} model fits we see that the value of $\alpha$ increases with radius,
while the width of the Gaussian DEM distribution shows a drop in the 1.0--2.0$\arcmin$ bin.
As expected from the EPIC profiles, the O/Fe ratio is lower in the core than in the outer parts.
The Ne/Fe ratio, however, is consistent with being flat within 2$\arcmin$ from the core. 
Finally, the width of the lines, indicated by the scale parameter (see Sect.~\ref{sec:rgs_analysis})
, increases outside the core. There it is consistent with being 1.0. In these models we 
fit the average of all the line widths with respect to the continuum surface brightness. 

From the O and Fe abundances measured with EPIC and RGS in the inner 2$\arcmin$ of the cluster,
we detect a jump in the O/Fe ratios between the inner 0.5$\arcmin$ and the annulus from 0.5--2.0$\arcmin$.
To illustrate this, we combine the single-temperature EPIC and RGS results
in the 0.5--2.0$\arcmin$ region. From EPIC we obtain (O/Fe)$_{0.5\arcmin-2.0\arcmin}$ = 1.14 $\pm$ 0.10 
which is significantly higher than the central (O/Fe)$_{0.0\arcmin-0.5\arcmin}$ = 0.76 $\pm$ 0.12. 
The combined O/Fe ratio from the 0.5--2.0$\arcmin$ RGS results is also significantly higher, 
(O/Fe)$_{0.5\arcmin-2.0\arcmin}$ = 0.84 $\pm$ 0.09, compared to the central value (O/Fe)$_{0.0\arcmin-0.5\arcmin}$ = 0.53 
$\pm$ 0.05. This jump has a confidence of 2.5$\sigma$ and 3$\sigma$ in EPIC and RGS respectively, which corresponds to a combined significance 
of 3.9$\sigma$. Despite the fact that the absolute values of the O/Fe ratio are different for the EPIC 
and RGS results, the relative increase of the O/Fe ratio is the same for both instruments: 1.5 $\pm$ 0.3 
(EPIC) and 1.6 $\pm$ 0.2 (RGS). In the {\it wgem} and {\it gdem} results the relative jump in O/Fe is equal
or lower, but still consistent with the result for the single temperature model: 1.4 $\pm$ 0.3 (EPIC/{\it wdem}),
1.4 $\pm$ 0.2 (EPIC/{\it gdem}), 1.6 $\pm$ 0.3 (RGS/{\it wdem}) and 1.5 $\pm$ 0.3 (RGS/{\it gdem}).
  
\begin{table}[tb]
\caption{Fit results for spatially resolved RGS spectra between 8--38 \AA~excluding CCD2. The fitted models are
single-temperature CIE (s), {\it wdem} (w) and a Gaussian DEM model (g). 
Emission measures ($Y = \int n_e n_H \mathrm{d}V$) are given in 10$^{66}$ cm$^{-3}$.
The iron abundance is fixed to 1.0 with respect to solar abundances.}
\begin{center}
\begin{tabular}{l|c|ccc}
\hline\hline
Parameter	& Mod		& 0--0.5$\arcmin$	& 0.5--1.0$\arcmin$	& 1.0--2.0$\arcmin$ \\
\hline
$Y$		& s		& 3.69$\pm$0.04		& 1.80$\pm$0.04		& 1.46$\pm$0.03	\\
		& w		& 3.60$\pm$0.04		& 1.78$\pm$0.04		& 1.30$\pm$0.05	\\
		& g		& 3.74$\pm$0.04		& 1.88$\pm$0.05		& 1.46$\pm$0.04	\\
\hline
$kT$ 		& s		& 3.23$\pm$0.09		& 4.2$\pm$0.2		& 4.3$\pm$0.3 \\
(keV)		& g		& 4.09$\pm$0.16		& 5.1$\pm$0.6		& 4.3$\pm$0.3 \\
\hline
$kT_{\mathrm{mean}}$& w		& 3.53$\pm$0.14		& 4.7$\pm$0.4		& 4.3$\pm$0.5 \\
\hline
$kT_{\mathrm{max}}$ & w		& 4.76$\pm$0.18		& 6.6$\pm$0.5		& 6.5$\pm$0.7 \\
\hline
$\alpha$	& w		& 0.53$\pm$0.04 	& 0.66$\pm$0.09		& 1.06$\pm$0.18 \\
\hline
$\sigma_{\mathrm{T}}$	& g	& 0.342$\pm$0.018	& 0.35$\pm$0.06		& 0.00$\pm$0.12 \\
\hline
O/Fe		& s		& 0.53$\pm$0.05		& 0.77$\pm$0.12		& 0.96$\pm$0.15 \\
		& w		& 0.51$\pm$0.05		& 0.72$\pm$0.12		& 1.2$\pm$0.3 \\
		& g		& 0.54$\pm$0.05		& 0.68$\pm$0.18		& 0.93$\pm$0.17 \\
\hline
Ne/Fe		& s 		& 1.10$\pm$0.12		& 1.7$\pm$0.3		& 0.8$\pm$0.4 \\
		& w		& 0.79$\pm$0.11		& 0.9$\pm$0.3		& 0.00$\pm$0.10 \\
		& g 		& 0.75$\pm$0.12		& 0.9$\pm$0.4		& 0.8$\pm$0.4 \\
\hline
Mg/Fe		& s		& 0.16$\pm$0.10		& 0.00$\pm$0.10		& 0.00$\pm$0.04 \\
		& w		& 0.32$\pm$0.11		& 0.0$\pm$0.2		& 0.00$\pm$0.12 \\
		& g		& 0.28$\pm$0.11		& 0.00$\pm$0.11		& 0.00$\pm$0.05 \\
\hline
Scale		& s		& 0.50$\pm$0.08		& 1.14$\pm$0.17		& 1.3$\pm$0.3 \\
		& w		& 0.45$\pm$0.07		& 1.1$\pm$0.2		& 2.8$\pm$0.4 \\
		& g		& 0.40$\pm$0.07		& 0.8$\pm$0.3		& 1.3$\pm$0.2 \\		
\hline
$\chi^2$ / d.o.f.& s		& 1100 / 858		& 970 / 858		& 925 / 858	\\
		& w		& 1000 / 855		& 953 / 855		& 885 / 855	\\
		& g		& 1039 / 857		& 971 / 857		& 926 / 857	\\
\hline
\end{tabular}
\label{tab:rgs_sp}
\end{center}
\end{table}

\subsection{Abundances and SNIa/SNII/Population-III models}

From the single-temperature (CIE) and DEM models we fit to the data, we
obtain the abundances of the elements for which line emission is detected. 
Assuming that all the elements originate from SNIa, SNII and  PopIII stars, we can construct
a simple model to obtain the relative contribution of these objects to the enrichment of the ICM. 

We use several SNIa yields obtained from two physically different models adapted from 
\citet{iwamoto1999} to fit our abundances. The W7 model describes a so-called slow deflagration model, while 
the WDD2 is calculated using delayed-detonation (DD) models, which is the currently favoured Type Ia 
explosion scenario.  
Note that with SNII we mean all types of core-collapse supernovae including types
Ib and Ic. We use the SNII yields integrated over the stellar population calculated by \citet{tsujimoto1995}
and \citet{iwamoto1999}. For the PopIII-star SN yields we use two models from \citet{heger2002} with  
different core masses of the PopIII star: 65 M$_{\sun}$ and 130 M$_{\sun}$. These two masses are
the lowest and the highest core mass considered in \citet{heger2002}.

For every element $i$ the total number of particles $N_i$ is a linear combination of the number 
of atoms produced by supernova type Ia (Y$_{i,\mathrm{Ia}}$), type II (Y$_{i,\mathrm{II}}$) and PopIII stars 
(Y$_{i,\mathrm{III}}$). 
\begin{equation}
N_i = a\mathrm{Y}_{i,\mathrm{Ia}} + b\mathrm{Y}_{i,\mathrm{II}} + c\mathrm{Y}_{i,\mathrm{III}},
\end{equation}
where $a$, $b$ and $c$ are multiplication factors of SNIa, SNII, and PopIII stars respectively. 
The total number of particles for an element can be easily converted into a number abundance.
This reduces to a system of three variables ($a$, $b$ an $c$) and nine data points 
(O, Ne, Mg, Si, S, Ar, Ca, Fe and Ni). We present the ratio of the relative numbers 
of SNIa, SNII and PopIII with respect to the total number.

\begin{table}
\caption{Relative contribution of SNIa, SNII and PopIII stars to the enrichment of the ICM. We compare
two SNIa models in \citet{iwamoto1999} with the data. We only show results from {\it wdem} fits,
because we could only get an acceptable fit using these data. The values shown here are the fractions
with respect to the sum of all contributions (SNIa + SNII + PopIII).}
\begin{center}
\begin{tabular}{l|cc|cc}
\hline\hline
Type	& \multicolumn{2}{c}{$M_{\mathrm{PopIII}}=130 $M$_{\sun}$}	& \multicolumn{2}{c}{$M_{\mathrm{PopIII}}=65 $M$_{\sun}$} \\
\hline
\hline
	& Value		& $\chi^2$ / dof&  Value	& $\chi^2$ / dof      \\
\hline
\multicolumn{5}{c}{W7} \\
\hline
SNIa  & 0.33$\pm$0.09 & 	      & 0.30$\pm$0.04 & 		    \\
SNII  & 0.7$\pm$0.2   & 41 / 6        & 0.70$\pm$0.14 & 55 / 6  	    \\
PopIII&(3.4$\pm$1.4)$\times$10$^{-3}$&&(-3$\pm$7)$\times$10$^{-3}$ &	    \\
\hline
\multicolumn{5}{c}{WDD2} \\
\hline
SNIa  & 0.38$\pm$0.08 & 	      & 0.48$\pm$0.10 & 		    \\
SNII  & 0.63$\pm$0.15 & 12 / 6        & 0.5$\pm$0.2   & 12 / 6  	    \\
PopIII&(-1.8$\pm$1.2)$\times$10$^{-3}$&&(1.7$\pm$1.1)$\times$10$^{-2}$ &     \\
\hline 
\end{tabular}
\vspace{0.3cm}\\

{\it Fits with the calcium abundance excluded:}
\begin{tabular}{l|cc|cc}
\hline\hline
Type	& \multicolumn{2}{c}{$M_{\mathrm{PopIII}}=130 $M$_{\sun}$}	& \multicolumn{2}{c}{$M_{\mathrm{PopIII}}=65 $M$_{\sun}$} \\
\hline
\hline
	& Value		& $\chi^2$ / dof&  Value	& $\chi^2$ / dof      \\
\hline
\multicolumn{5}{c}{W7} \\
\hline
SNIa  & 0.34$\pm$0.09 & 	      & 0.32$\pm$0.05 & 		    \\
SNII  & 0.7$\pm$0.2   & 29 / 5        & 0.69$\pm$0.15 & 41 / 5  	    \\
PopIII&(2.9$\pm$1.3)$\times$10$^{-3}$&&(-2$\pm$7)$\times$10$^{-3}$ &	    \\
\hline
\multicolumn{5}{c}{WDD2} \\
\hline
SNIa  & 0.38$\pm$0.08 & 	      & 0.50$\pm$0.10 & 		    \\
SNII  & 0.62$\pm$0.15 & 4.9 / 5        & 0.5$\pm$0.2   & 4.1 / 5  	    \\
PopIII&(-1.9$\pm$1.2)$\times$10$^{-3}$&&(1.9$\pm$1.2)$\times$10$^{-2}$ &     \\
\hline 
\end{tabular}
\end{center}
\label{tab:abunfit}
\end{table}

We fit the abundances we obtain from the EPIC results of the core region to a model consisting of SNIa, SNII 
and PopIII-star yields in order to determine their relative contribution to the ICM.
In Table~\ref{tab:abunfit} we compare two SNIa yield models  fitted
together with the SNII and PopIII yield models to the measured abundances in 
Table~\ref{tab:4arcmin}. We use the abundances derived from the fits with the {\it wdem} model. 
If we include the calcium abundance, none of the fits are statistically acceptable. However, if we
ignore the calcium data point, we obtain a $\chi^2_r$ of about 1 for the WDD2 model. 
We also test other SNIa models listed in \citet{iwamoto1999}:
W70, WDD1, WDD3, CDD1 and CDD2. Their best-fit ratios were in general similar to the WDD2 and W7 
models, but with a higher $\chi^2$.

\begin{figure*}[t]
\begin{minipage}{0.5\textwidth}
\includegraphics[width=\columnwidth]{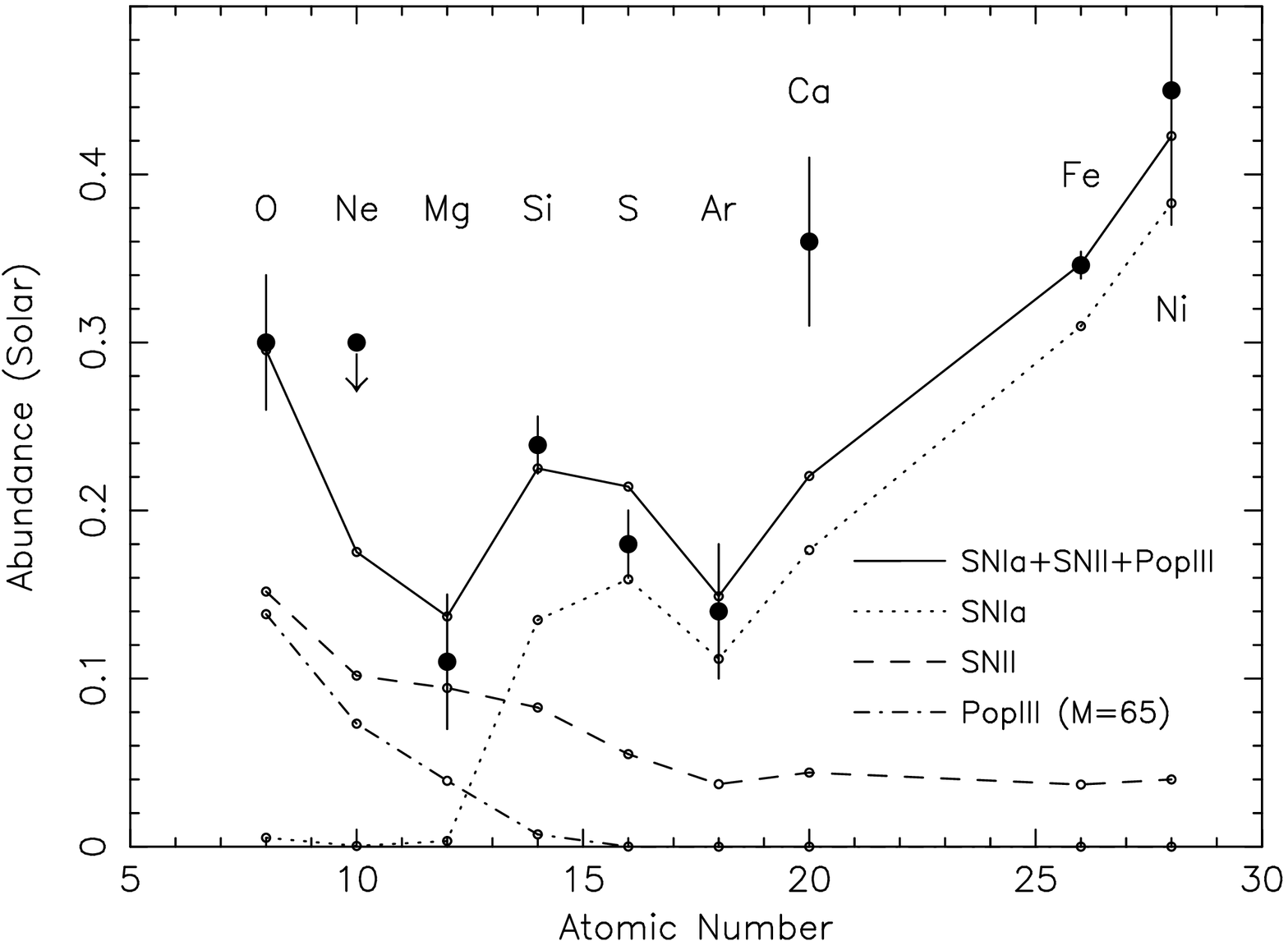}
\end{minipage}
\begin{minipage}{0.5\textwidth}
\includegraphics[width=\columnwidth]{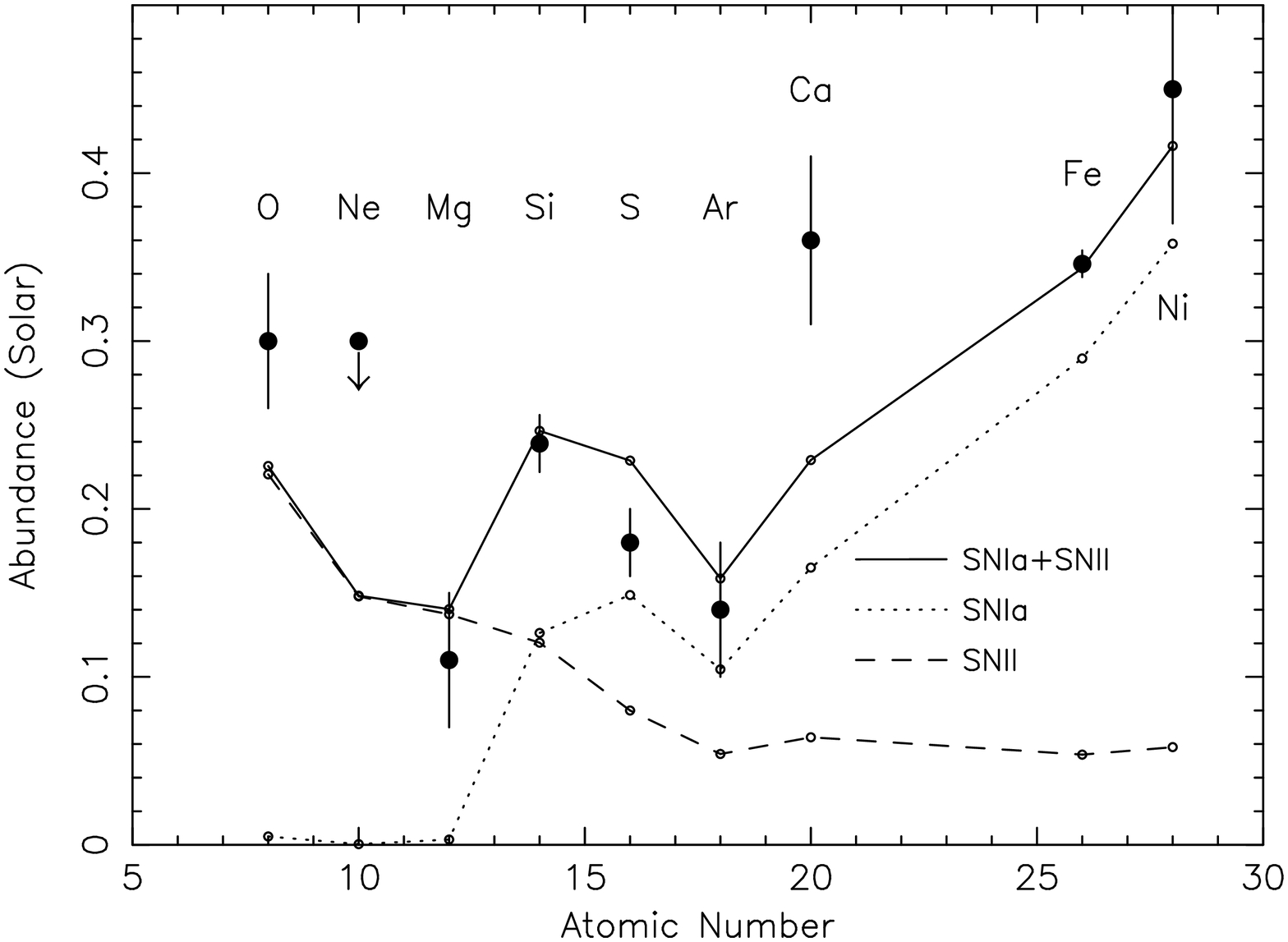}
\end{minipage}
\caption{The elemental abundances derived from a {\it wdem} fit, fitted with the WDD2 SNIa-yield model.
The best fit from Table~\ref{tab:abunfit} is shown in the left panel with a PopIII mass of 65 $M_{\sun}$. In the plot
we also show the SNIa, SNII and PopIII contribution separately. In the right panel we show a fit without
a PopIII contribution. In both plots the calcium abundance is included in the fit. }
\label{fig:abun_wdem}
\end{figure*}

The relative contribution of PopIII stars is in all fits smaller than three times its error, 
thus it is not significantly detected. Because of the large yields per PopIII event, a small number 
of PopIII stars can in principal contribute a lot to the abundance. The abundance pattern of PopIII stars, however, 
resemble the patterns of SNIa and SNII depending on the core mass. The SNIa or SNII patterns can mostly 
compensate for the PopIII contribution, when it is left out in the fit. The $\chi^2$ does improve 
only marginally when we add PopIII star yields to a model containing only SNIa and SNII yields. The best fit 
using just SNIa and SNII models including calcium gives a $\chi^2$/dof of 15 / 7 and a SNIa contribution 
of 0.35$\pm$0.03. In fact some models fit the data best by putting a negative value to the PopIII or 
SNII contribution, which is unphysical. The 65 M$_{\sun}$ PopIII
model produces a slightly higher PopIII contribution than the one for 130 M$_{\sun}$. But since the mass 
of the expelled material is higher for the latter model, the number of stars needs to be higher in the
65 M$_{\sun}$ model to get a comparable effect. The errors on these bigger values are also too large 
to claim a significant PopIII-star contribution.

In Fig.~\ref{fig:abun_wdem} we plot the abundances from the {\it wdem} model with the best-fit model
(WDD2) listed in Table~\ref{tab:abunfit}. Note that the WDD2 model is also the type Ia model favoured
by \citet{iwamoto1999} on observational grounds. The data points we use are taken from the third
column of Table~\ref{tab:4arcmin}. From this plot we see that the calcium 
abundance might be underestimated by the nucleosynthesis models and the cause of the high $\chi^2$ values.
But the actual uncertainties in the measured calcium abundance might be bigger than the statistical error indicated here. 
This calcium overabundance is also seen in the analysis of 2A 0335+096 \citep{werner2005}, confirming that
the overabundance is probably not a statistical deviation. 
Every other abundance is consistent with the model within the error bars. The contribution of PopIII
can be seen mainly in oxygen and neon, but the uncertainties are large.

We applied the same procedure to the abundances obtained from RGS. The relative abundances determined from RGS 
are very well constrained, but the number of elements that we can measure is too small to get a reasonable fit. 
If we fit the abundances we obtain errors which are equal to the measured values. Therefore, we do not show 
these fits in this paper.

\subsection{Non-thermal X-ray emission?}
\label{sec:hardsoft}

\begin{figure*}
\includegraphics[width=\textwidth]{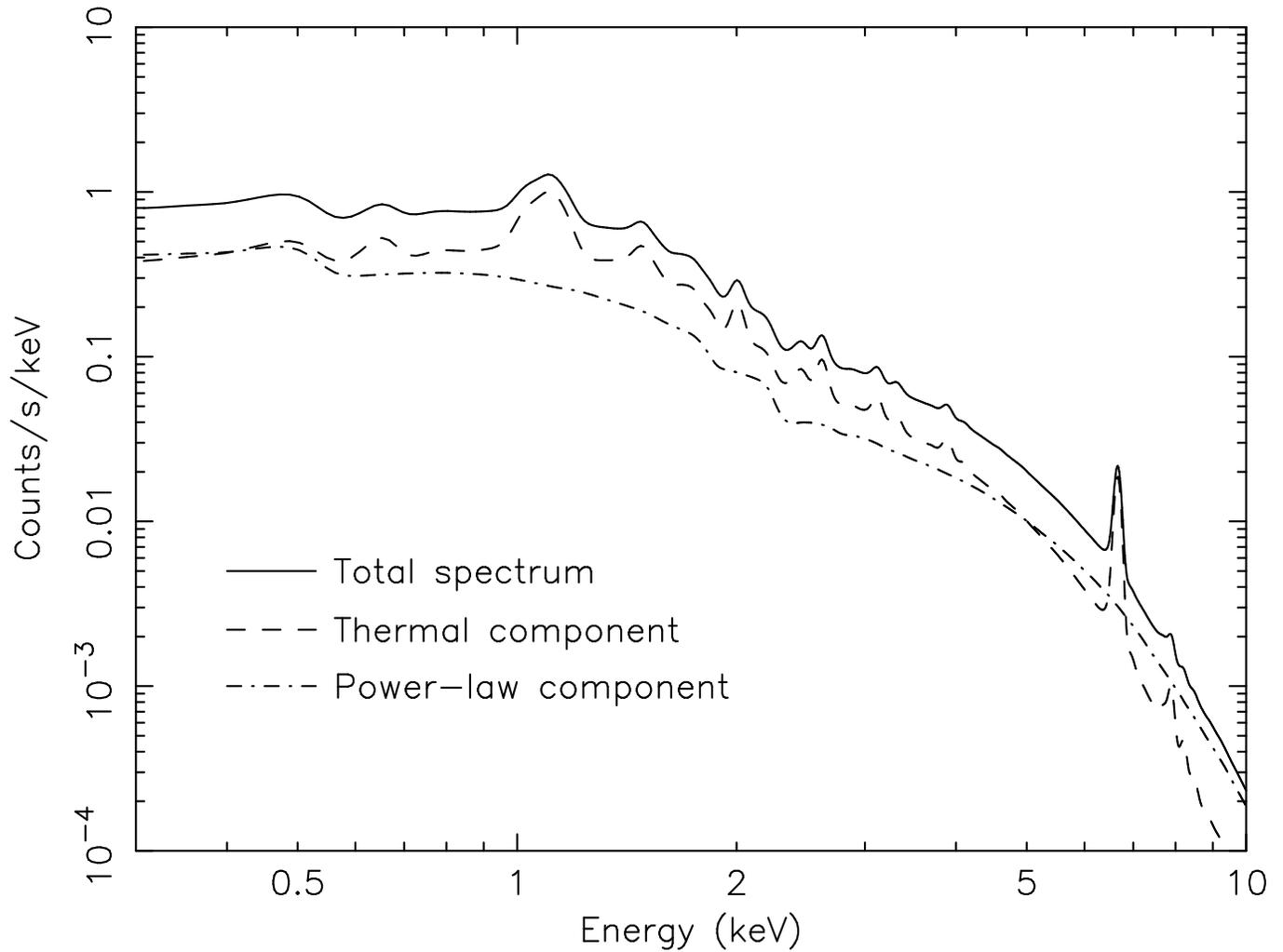}
\caption{This spectrum shows that an underlying power-law component can cause both a soft and a hard excess.
The figure shows a power-law component with $\Gamma$ = 2, a single-temperature CIE component with
a temperature of 2.5 keV and the total spectrum. The contribution of the power-law component is 
mostly noticeable below $\sim$0.5 keV and above $\sim$5 keV, hence causing a soft and hard excess
with respect to the thermal component.}
\label{fig:softhard}
\end{figure*}

\begin{figure*}[!tb]
\begin{minipage}{0.5\textwidth}
\includegraphics[width=1.0\textwidth]{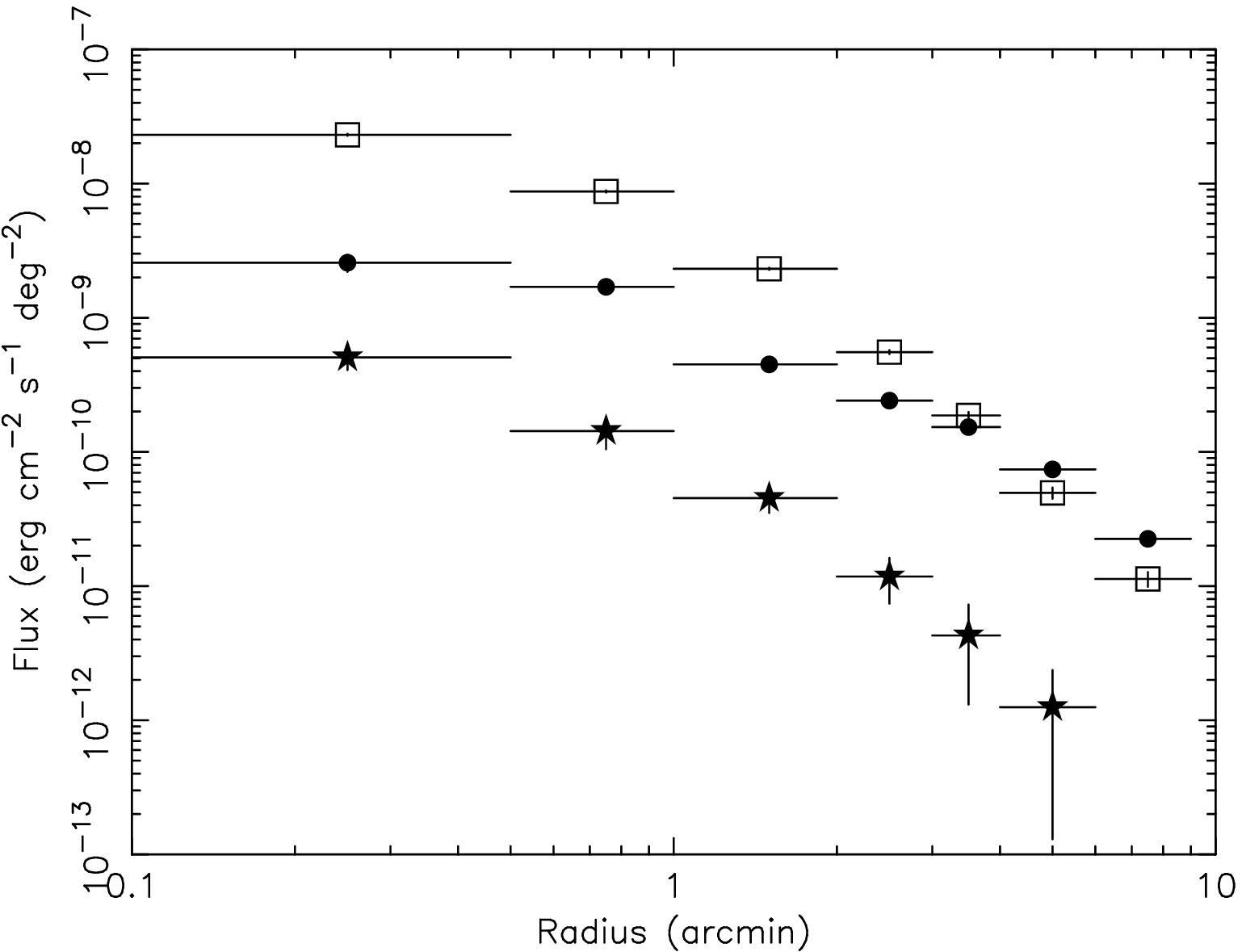}
\end{minipage}
\begin{minipage}{0.5\textwidth}
\includegraphics[width=1.0\textwidth]{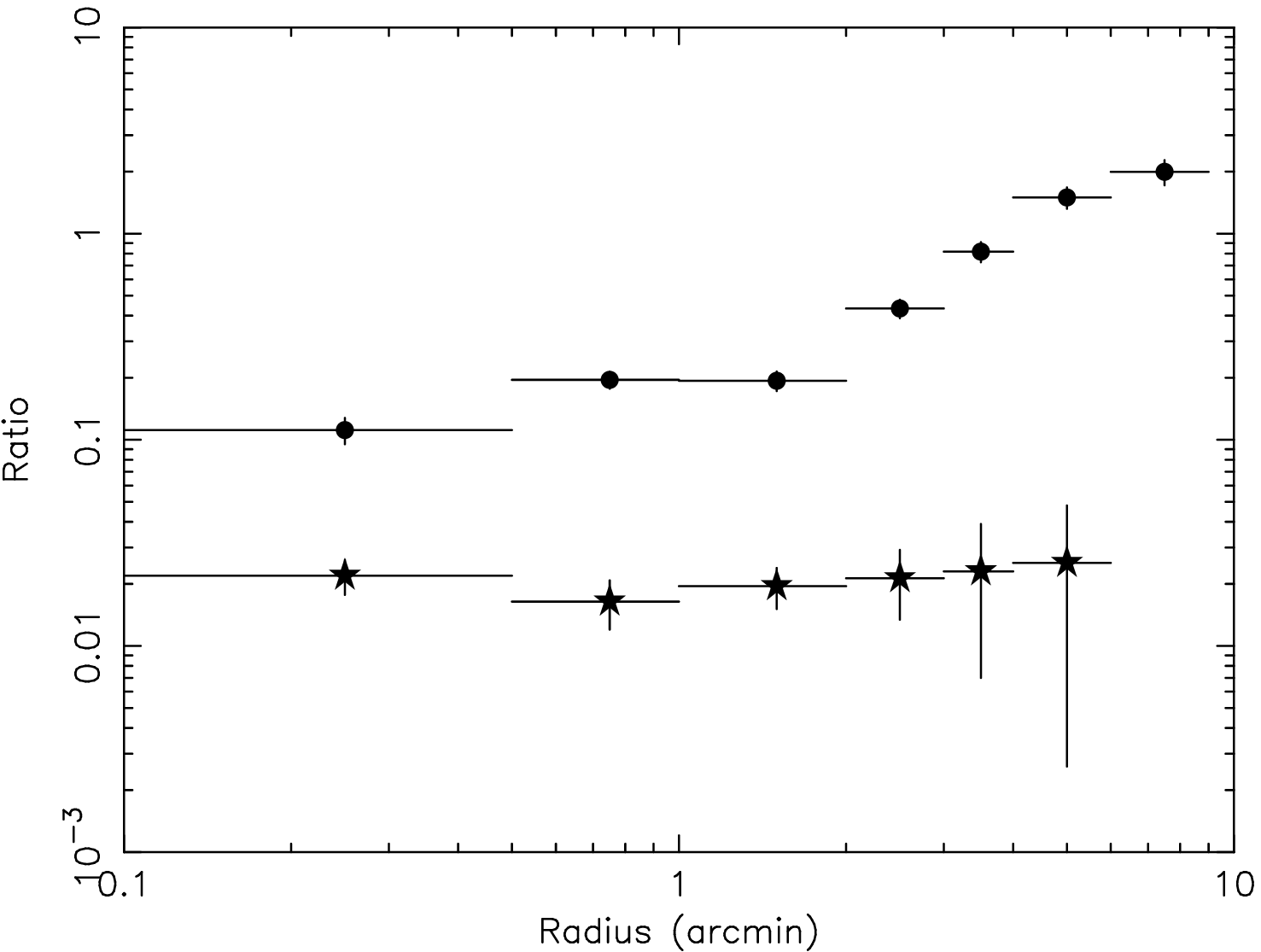}
\end{minipage}\\
\caption{Left: The 0.3--10 keV integrated intensities for three model components: Cluster hot gas ($\Box$), 
power law ($\bullet$) and soft-excess ($\star$). Right: Ratio between the intensities: power law/hot cluster gas
($\bullet$) and soft-excess/hot cluster gas ($\star$).}
\label{fig:epic_norms}
\end{figure*}

In previous papers about S\'ersic 159-03 \citep[e.g.][]{kaastra2003a,bonamente2005} the larger $\chi_r^2$ for the 
single-temperature model is attributed to a soft X-ray excess. There are, however, a number 
of solutions to fit this soft excess and to obtain an acceptable $\chi_r^2$ of about 1.0. 
From the spectral fits in Sect.~\ref{sec:epicrgs} we confirm that a single-temperature model is not 
the best description for the observed spectra. If we fit the data with a multi-temperature model, then 
we obtain a $\chi_r^2$ of about 1.0. In this section we explore the possibility of the existence of
a soft excess in S\'ersic 159-03 and speculate about its nature. The {\it wdem} and {\it gdem} 
models already provide acceptable fits to the spectra. We verified that the non-thermal component 
presented in this section does not affect the trends observed in the thermal analysis listed 
in Sect.~\ref{sec:epicrgs}. 

One possibility is that the soft X-ray emission originates from warm thermal emission 
from the WHIM \citep{kaastra2003a}.
Here we explore the possibility that inverse-Compton emission from CMB photons which 
are up-scattered to X-ray energies by relativistic electrons can also cause a soft excess. 
The same emission mechanism is also thought to be the origin of the hard excess in clusters 
of galaxies detected by BeppoSAX \citep[e.g.][]{fusco-femiano2005}. A power-law component 
describing this inverse-Compton emission might be able to fit both this hard and soft excess.

A single non-thermal component as illustrated in Fig.~\ref{fig:softhard} with a $\Gamma$ of 2.0
can explain both hard and soft X-ray excess with respect to a single-temperature model.
At low energies $\lesssim$0.5 keV the power-law component
is comparable in flux to the thermal component, hence causing a soft excess. Above $\sim$5 
keV the power-law component in this example is even stronger than the thermal component, causing a hard X-ray excess.
In S\'ersic 159-03 the power-law emission is not as strong as in this example, but a similar model
is consistent with the data.

\begin{table}
\caption{Fit results for an EPIC spectrum extracted from a circle with a radius of 4$\arcmin$ 
and centred on the core including a single-temperature thermal component + power law and multi-temperature
thermal components + power law.
Fluxes are calculated over the 0.3--10 keV range and presented in 10$^{-10}$ erg cm$^{-2}$ s$^{-1}$ deg$^{-2}$.
Emission measure ($Y_{\mathrm{thermal}} = \int n_e n_H \mathrm{d}V$)
is given in 10$^{66}$ cm$^{-3}$ and $Y_{\mathrm{pow}}$ is given in 10$^{51}$ ph s$^{-1}$ keV$^{-1}$ at 1 keV.}
\begin{center}
\begin{tabular}{l|c|c|c}
\hline\hline
Parameter	& single-temp + pow 	& {\it wdem} + pow	& {\it gdem} + pow \\	
\hline
$Y_{\mathrm{thermal}}$
		& 16.8 $\pm$ 0.3	& 17.6 $\pm$ 0.3	& 18.9 $\pm$ 0.4	\\
$F_{\mathrm{thermal}}$
		& 16.2 $\pm$ 0.2 	& 16.9 $\pm$ 0.3	& 18.0 $\pm$ 0.5	\\
$kT$		& 2.45 $\pm$ 0.02	& 			& 2.50 $\pm$ 0.03	\\
$kT_{\mathrm{mean}}$&			& 2.53 $\pm$ 0.07	&		\\
$kT_{\mathrm{max}}$& 			& 3.14 $\pm$ 0.06	& 	\\
$\alpha$	&			& 0.32 $\pm$ 0.04	& 	\\
$\sigma_{T}$	&			& 			& 0.208 $\pm$ 0.012	\\
O		& 0.27 $\pm$ 0.05	& 0.26 $\pm$ 0.05	& 0.17 $\pm$ 0.03	\\
Ne		& 1.00 $\pm$ 0.11	& 0.42 $\pm$ 0.11	& 0.11 $\pm$ 0.09	\\
Mg		& 0.18 $\pm$ 0.04	& 0.21 $\pm$ 0.05	& 0.18 $\pm$ 0.04	\\
Si		& 0.32 $\pm$ 0.02	& 0.33 $\pm$ 0.02	& 0.33 $\pm$ 0.02	\\
S		& 0.22 $\pm$ 0.02	& 0.25 $\pm$ 0.02	& 0.25 $\pm$ 0.02	\\
Ar		& 0.20 $\pm$ 0.05	& 0.23 $\pm$ 0.05	& 0.26 $\pm$ 0.06	\\
Ca 		& 0.42 $\pm$ 0.06	& 0.45 $\pm$ 0.06	& 0.52 $\pm$ 0.07	\\
Fe		& 0.433 $\pm$ 0.011	& 0.409$\pm$ 0.011	& 0.275 $\pm$ 0.009	\\
Ni		& 0.62 $\pm$ 0.11	& 0.60 $\pm$ 0.10	& 0.54 $\pm$ 0.10 	\\
$Y_{\mathrm{pow}}$
		& 6.6 $\pm$ 0.4		& 5.3 $\pm$ 0.4		& 3.5 $\pm$ 0.8		\\
$F_{\mathrm{pow}}$
		& 3.5 $\pm$ 0.2		& 2.8 $\pm$ 0.2		& 1.6 $\pm$ 0.4	\\
$\Gamma$	& 2.06 $\pm$ 0.03	& 2.10 $\pm$ 0.04	& 2.30 $\pm$ 0.15	\\
\hline
$\chi^2$ / dof	& 897 / 914		&  853 / 913		 & 861 / 913		 \\	 
\hline
\end{tabular}
\label{tab:4arcmin_powerlaw}
\end{center}
\end{table}

In order to learn whether the data allow the soft excess to be thermal or non-thermal in nature,
we fit the spectra from all annuli (see Table~\ref{tab:annuli}) with a single-temperature 
cluster component, a cold thermal component representing the WHIM, and a power-law component 
representing the soft and hard non-thermal emission. We first fit all annuli with the photon index 
and the $kT$ of the cool component as a free parameter. Because the photon index and temperatures 
were not well constrained in the outer annuli, we fix the photon index and cool temperature 
in the final fits of all annuli to the mean value in the core region, which is 1.9 for the power-law 
index and 0.25 keV for the cool thermal component. This assumption for $\Gamma$ is valid if the relativistic 
electrons are produced by acceleration in shocks, which behave similarly regardless of the 
position in the cluster. The radial profiles of the fitted components are shown in 
Fig.~\ref{fig:epic_norms}. The profile of the hot gas and the soft component have 
the same shape, which suggests that they have the same origin. The power-law component,
however, has a more extended radial distribution. The flux is still the highest in the core
of the cluster, but the slope of the radial profile is more shallow than the thermal emission. 
The ratio between the components is plotted in the right panel of Fig.~\ref{fig:epic_norms}. 
This figure indeed shows that the ratio between the soft excess and the cluster gas is constant. 
This is is not what you expect if the excess is due to WHIM which should be picked up preferably
at the outskirts of the cluster. On the other hand, the power-law component does become 
relatively more important in the outskirts of the cluster. This is to be expected for 
Inverse-Compton emission because it scales with density $n$ instead of $n^2$ which 
is true for thermal emission. The different profile of the power-law emission therefore
strongly suggests that if there is an excess, it is non-thermal in nature.

In the following fits we drop the cool component and fit the spectra with just a thermal (DEM) components and 
an additional power-law component to fit the soft-excess (Table~\ref{tab:4arcmin_powerlaw}).
The models including a power law generally result in a lower $\chi^2$. This effect is most strongly present 
in the single-temperature fit: $\chi^2$ = 897 / 914 with a power-law component and $\chi^2$ = 1228 / 916 
without. But also for the DEM models there is an improvement in $\chi^2$ that provides an argument for 
the presence of a non-thermal component.

The absolute values of the abundances change when an additional power law is used, but the general
trends do not change considerably. The spatial abundance distributions are also consistent with the trends 
observed in the single-temperature models. In particular, fits including a power-law component show that 
the jump in O/Fe in the centre of the cluster remains. However, the significance of the jump decreases 
compared to the single-temperature result when a power-law component is added.

Our suggestion that a single power-law component can both explain a possible soft as well as a hard excess 
of S\'ersic 159-03, may also be relevant for several clusters for which non-thermal emission has been detected by
BeppoSAX \citep{fusco-femiano2000,fusco-femiano2004,fusco-femiano2005}. However, the Coma-cluster is, so-far,
the only cluster for which both a hard X-ray component and a soft excess has been reported 
\citep{fusco-femiano2004,kaastra2003a}. For  S\'ersic 159-03, it does indeed seem to be the case
that a single non-thermal component explains both the soft and hard X-ray excess: Extrapolating the BeppoSAX 
flux in the 20--80~keV range, (1.5$\pm$0.5)$\times$10$^{-11}$~erg s$^{-1}$ cm$^{-2}$, to the 0.3--10 keV band, 
assuming  $\Gamma =2.0$, we obtain a flux of (3.8$\pm$1.3)$\times$10$^{-11}$~erg s$^{-1}$ cm$^{-2}$. This is 
consistent with the XMM-Newton flux for the putative non-thermal component reported by \citet{kaastra2003a},
(6.8$\pm$2.0)$\times$10$^{-11}$ erg s$^{-1}$ cm$^{-2}$. For this calculation we assume that all the non-thermal
emission originates from within a circle of 12$\arcmin$ from the core.

\section{Discussion}
\label{sec:discussion}

\subsection{Temperature structure}

The parameters of the DEM distributions which we obtain from multi-temperature fitting, are well 
determined for this cluster. However,
we cannot discriminate between different shapes of DEM distributions and the presence of 
a non-thermal component. The DEM parameters $\alpha$ ({\it wdem}) 
and $\sigma_{\mathrm{T}}$ ({\it gdem}) values show a slight increase to the outer parts of the cluster. 
This can be explained by the steep temperature gradients we see both in the cooling core
and in the outer parts of the cluster. The values we derive for $\alpha$
are higher than the value of 0.20 $\pm$ 0.05 derived by \citet{kaastra2004} based 
on a shorter exposure. However, their background subtraction method and handling 
of spectral excess is different, which can lead to systematic differences. Also projection 
effects can have an influence on the broadness of the DEM distribution, but as 
\citet{kaastra2004} and \citet{werner2005} show, multi-temperature models are also 
needed for fitting deprojected spectra.

\subsection{Abundance distribution and enrichment by supernova types Ia/II and Population-III stars}

From the EPIC and RGS spectra we obtain radial abundance profiles for the most abundant
metals. The EPIC and RGS radial profiles for oxygen show a jump in the centre of the cluster. 
The key data point is the point in the 0--0.5$\arcmin$ bin. If we compare the O/Fe ratio of the central
bin with the neighbouring bins, its value is significantly lower. This means that either the
iron abundance is relatively high or oxygen is low in the centre of the cluster.  

Recently, using cluster simulations, \citet{schindler2005} found
that ram-pressure stripping acting on in-falling galaxies can result in centrally peaked 
abundance profiles, while early galactic winds produce an extended distribution
of the oxygen abundance. The spatial distribution of oxygen, which is difficult 
to determine, is the key to understand the evolution history of the cluster.

Contrary to the iron and oxygen lines in S\'ersic 159-03, the widths of these 
lines are the same in the cluster 2A 0335+096 \citep{werner2005}. A possible explanation
is that the cluster 2A 0335+096 shows a much stronger temperature drop in the centre of the
cluster. Since spectral lines are stronger when the temperature is lower, the RGS spectrum is
dominated by line emission from the cool core of the cluster. In this case, the line profiles 
of iron and oxygen should follow the temperature structure more than the abundance distribution 
of the elements. In S\'ersic 159-03 the temperature profile within the RGS extraction region is relatively
flat. The temperature is about 2.5 keV with a spread of 10\%. Because a strong thermal
gradient is absent in the core of this cluster, the line profile should follow the abundance distribution
of the elements. The observed line profiles in RGS therefore strongly suggest that in S\'ersic 159-03 
oxygen emission is lower in the centre of the line profile compared to iron, which is consistent with the
radial profiles extracted from EPIC and RGS. 

From several abundance yield simulations for Supernova type Ia and II we know that oxygen 
dominates the yield of type II supernovae, while iron originates mostly from type Ia supernovae.
If the iron abundance is more peaked in the centre of the cluster, we can formulate a scenario of the 
possible enrichment history of the ICM in
\object{S\'ersic 159-03}. In the early universe the Inter-Galactic Medium (IGM) might have been enriched by  
PopIII-star explosions. The star bursts in the young galaxies that form after PopIII enrichment
produce a lot of SNII which produce mainly oxygen. Because clusters are still in an early stage
of development, the oxygen becomes well mixed through the IGM. 
Then, 1 billion years after the 'Big Bang', also SNIa explosions
start enriching the local ISM in galaxies continuously. If a galaxy containing both SNIa and SNII products 
falls into a cluster, the ISM is stripped off by ram-pressure stripping, preferably in the denser core of 
the cluster, hence leading to a more centrally peaked abundance distribution of all SNIa and SNII products 
\citep{schindler2005}. Because iron is also produced by SNIa explosions in the central cD galaxy, 
this could enhance the iron abundance in the centre of the cluster \citep{tamura2001b}. The 
drop in the centre of the observed O/Fe fits within this scenario.

In order to check whether the absolute abundance of iron can be consistent with the interpretation we just 
formulated, we calculate the total iron mass in the cluster. We find about 4 $\times$ 10$^9$ $M_{\sun}$ within
a radius of 300 kpc of the core. If all this iron originates from SNIa explosions, we would need at least
5 $\times$ 10$^9$ explosions in the cluster lifetime, which is of the order of 10$^{10}$ years. 
Since there are typically 10$^2$ galaxies, there should be about 1 SNIa explosion per 200 years per galaxy,
which is consistent with the rate of SNIa in our own galaxy.  

We then fit the abundances from EPIC to a model of the yields of SNIa, SNII and PopIII stars 
\citep{iwamoto1999,heger2002}. Using a linear combination of the yields we can estimate the expected 
abundances and fit them to the data. From the fit we obtain the relative contributions of the three types 
of metal-enriching sources. 

We have to be careful, though, when we interpret the fitted supernova ratios. 
The mechanisms by which the elements are ejected into the ICM are not yet fully understood and
are certainly not included in our simplified model. A small part of the supernova products can
also be locked up in low-mass stars, so the abundance distribution we observe in the ICM
is not entirely representative for the total abundance distribution in the cluster. A more detailed 
description of the problems when one uses this simplified model is described in \citet{matteucci2005}. 
The derived supernova ratios should be interpreted as the number fraction of supernovae
that would be needed to enrich the ICM, not the actual number of supernovae during the  
history of the cluster.

Our analysis shows that PopIII stars are not required to fit the data. The contributions from
this early generation of stars are probably too small to be detectable in current cluster abundance patterns.
We cannot confirm the results from a sample of clusters observed with ASCA that suggest that low mass PopIII 
stars are necessary to explain the observed silicon and sulfur abundance \citep{baumgartner2005}. Our 
analysis, which includes more elements with about the same error bars on the abundances as 
in \citet{baumgartner2005}, shows that a PopIII-star contribution is not detected in S\'ersic 159-03.
Considering the single-temperature spectral models used by \citet{baumgartner2005}, we think
that a temperature bias in the iron abundance makes their Si/Fe and S/Fe ratios unreliable. 
We also cannot confirm the underabundance of argon and calcium in the \citet{baumgartner2005} 
sample. We actually measure an overabundance of calcium with respect to the nucleosynthesis models, 
which is in line with the result by \citet{werner2005} 
in the cluster \object{2A 0335+096}. This might be an indication that the 
SNII and SNIa models have difficulties predicting the right yield for this element, but the actual uncertainties
in our measured value might also be bigger than the quoted statistical error.  
When we compare the supernova fractions found by \citet{werner2005} in \object{2A 0335+096}, 
we see that the SNIa contribution in S\'ersic 159-03 tends to be 5--25\% higher.

\subsection{Soft excess and non-thermal X-ray emission}

An alternative solution to obtain an acceptable fit for the spectra of S\'ersic 159-03 is to add
a soft-excess component to a single-temperature model. We explore the possibility that the 
deviations from the single-temperature model are due to a different component, like
the WHIM or inverse-Compton emission. 

A non-thermal component can explain both a soft and hard excesses for both S\'ersic 159-03 and 
Coma, is consistent with inverse-Compton emission of CMB photons on relativistic electrons \citep{hwang1997}. 
The alternative explanation, non-thermal bremsstrahlung emission from a non-thermal tail to the Maxwellian
energy distribution of the electrons, would only result in a hard X-ray excess.

There are several possible sources that can produce a population of relativistic 
electrons in a cluster that would explain the observed power-law emission.
The acceleration of electrons can occur, for example, in active galactic 
nuclei, supernovae and pulsars \citep{sarazin1998}. But also plasma waves 
or shocks in the ICM can accelerate electrons \citep[see][~for a discussion]{ensslin1999}.  
Using our results from S\'ersic 159-03, we propose a new scenario where
electrons are accelerated by the shock waves associated with in-falling galaxies.

We already discussed the possibility that the centrally peaked iron distribution might be due to
an efficient ram-pressure stripping of in-falling galaxies in the dense core of 
the cluster. Some fraction of these galaxies being stripped
would have moderate supersonic velocities. The galaxies will drive bow shocks 
of Mach numbers up to $\sim$ 3 in the intra-cluster gas that can accelerate 
electrons to relativistic energies.
To produce photons of $\sim$ 10 keV due to inverse-Compton
up-scattering of CMB photons, electrons with a Lorentz factor $\gamma \sim 
7 \times 10^3$ are required. The life-time of such an electron in the
cluster core is about 2.3$\times$10$^{12} \gamma^{-1}$ years (assuming that 
the magnetic field in the core is below 2$\mu$G and the energy densities
of optical and IR radiation field in the core are below 0.1 eV cm$^{-3}$).
It is worth noting here that the central galaxy in S\'ersic 159-03 is one of the 
brightest cD galaxies detected in IR \citep{hansen2000}, but the energy 
density of the IR photon field produced by the galaxy is below 0.1 eV 
cm$^{-3}$.

The electron acceleration time at a shock of velocity $\geq$ 1000 km s$^{-1}$ 
in the cluster core is typically much shorter than 3$\times$10$^8$ years. Thus, to 
provide a steady non-thermal X-ray emission the stripping galaxies
must cross the cluster core at a frequency $\nu$ of about 1 in 3$\times$10$^8$ 
years. The mechanical power, $L_{\mathrm{kin}}$, dissipated by the galactic shocks of 
velocity $v_{\mathrm{g}}$ and average radius $R_{\mathrm{g}}$ in the core region of diameter 
$D_{\mathrm{c}}$ and average number density $n_{\mathrm{c}}$ can be written as:
\begin{eqnarray}
L_{\mathrm{kin}} & \approx & 2\times10^{43} ~ \left(\frac{R_{\mathrm{g}}}{15~\mathrm{kpc}}\right)^2 
~\left(\frac{D_{\mathrm{c}}}{300~ 
\mathrm{kpc}}\right)~\left(\frac{v_{\mathrm{g}}}{10^3 ~\mathrm{km~s^{-1}}}\right)^2 \nonumber \\
&  & \left(\frac{n_{\mathrm{c}}}{3\times10^{-3} ~\mathrm{cm^{-3}}}\right)~
\left(\frac{\nu}{3 \times 10^{-9} ~\mathrm{yr^{-1}}}\right)~  \quad \mathrm{erg~s^{-1}}
\end{eqnarray}
For the parameters assumed above, galaxies with velocity $v_{\mathrm{g}} \sim$ 2000 km 
s$^{-1}$ would provide $L_{\mathrm{kin}} \sim 10^{44}$ erg s$^{-1}$.
A few percent of the power would be enough to maintain the relativistic 
electrons required to provide the 10 keV regime non-thermal emission. Note that 
the lifetime of the electrons with Lorentz factor below 100
would be about the Hubble time. This implies that the spectrum of electrons 
accelerated by multiple successive shocks of Mach number about 3 will have 
power-law distribution of index about 3 at $\gamma \sim$ 10$^4$, and 
a flat distribution at $\gamma \leq$ 100, where the Compton
losses are not important. The electron distribution index 3 corresponds to the 
photon index of the photon emission of 2, close to that observed.

\section{Conclusions}

We have analysed high-resolution X-ray spectra of the cluster of galaxies S\'ersic 159-03 
obtained with XMM-Newton and conclude that:

\begin{enumerate}

\item For the first time we accurately determine the abundances in a cluster of galaxies which 
fit to a linear combination of current supernova yield models.

\item From the line width in RGS and the radial profiles from EPIC/RGS we find a jump in the O/Fe ratio 
around a radius of 0.5$\arcmin$ from the cluster centre. The O/Fe ratio in the centre of the cluster
is lower compared to its immediate surroundings. A combination of ram-pressure stripping \citep{schindler2005} 
and enrichment by SNIa in the central elliptical galaxy \citep{tamura2001b} can explain the observed profile.

\item We do not detect a contribution of PopIII stars in S\'ersic 159-03. This result is not in line with
the claim by \citet{baumgartner2005} that PopIII stars are necessary to explain the abundances measured
in an ASCA sample of clusters. The number contribution of SNIa 
with respect to the total number of supernovae we find, based on the measured abundances, is about 25--50\%.

\item The spectra can also be fitted with an additional non-thermal component with a power-law index of about 2.1 
on top of the thermal emission. The $\chi^2_r$ of these fits is comparable to the $\chi^2_r$ using only multi-thermal 
models. If a non-thermal component is present in S\'ersic 159-03, it can be explained by Inverse-Compton 
scattering of CMB photons and relativistic electrons.

\end{enumerate}

\begin{acknowledgements}
We would like to thank the anonymous referee for providing comments that helped to improve the presentation
of our results and conclusions in this paper. 
This work is based on observations obtained with XMM-Newton, an ESA science mission with instruments
and contributions directly funded by ESA member states and the USA (NASA). The Netherlands Institute
for Space Research (SRON) is supported financially by NWO, the Netherlands Organisation for Scientific 
Research.
\end{acknowledgements}

\bibliographystyle{aa}
\bibliography{clusters}

\end{document}